\documentclass[
reprint,
 amsmath,amssymb,
 aps,
prd
]{revtex4-1}

\usepackage{amssymb,amsmath,amsfonts}
\usepackage[dvipsnames]{xcolor}
\usepackage{graphicx}
\usepackage{verbatim}
\usepackage{color}
\usepackage[mathscr]{euscript}
\usepackage{framed}
\usepackage{mdframed}
\usepackage{amsmath}
\usepackage{amsfonts}
\usepackage{mathrsfs}
\usepackage{amssymb}
\usepackage{mathrsfs}  
\usepackage{subcaption}
\usepackage{caption}
\DeclareCaptionJustification{justified}{\leftskip=0pt \rightskip=0pt \parfillskip=0pt plus 1fil}
\usepackage{graphicx}
\usepackage{color}
\usepackage{slashed}
\usepackage{hyperref}
\usepackage{feyn}
\hypersetup{colorlinks, citecolor=bluscuro, linkcolor=bluscuro, urlcolor=bluscuro}
\definecolor{rossos}{cmyk}{0,1,1,0.55}
\definecolor{bluscuro}{rgb}{0.15, 0.2, .85}
\definecolor{bluchiaro}{cmyk}{1,.3,0.,0.1}

\graphicspath{{./Figures/}}

\usepackage{dcolumn}% Align table columns on decimal point
\usepackage{bm}
\usepackage{hyperref}
\usepackage[mathlines]{lineno}

\numberwithin{equation}{section}

\usepackage{comment}

\begin{document}
\title{PBH remnants  as dark matter produced in thermal, matter and runaway-quintessence post-inflationary scenarios}

\author{Ioannis Dalianis$^1$}\email{dalianis@mail.ntua.gr}
\affiliation{{\it {$^1$\, Physics Division, National Technical University of Athens\\ 15780 Zografou Campus, Athens, Greece}}}
\author{George Tringas$^1$}\email{tringas@uni-bonn.de}
\affiliation{{\it {$^1$\, Physics Division, National Technical University of Athens\\ 15780 Zografou Campus, Athens, Greece}}}

\date{\today}

\begin{abstract}
We investigate the cosmology of mini Primordial Black Holes (PBHs) produced by large density perturbations. 
The mini PBHs evaporate promptly in the early universe and we assume that a stable remnant is left behind. The PBHs remnants can constitute the entire dark matter of the universe for a wide range of remnant masses.
We build inflationary models, in the framework of $\alpha$-attractors utilizing exponential functions,  in which the PBHs are produced during matter, radiation and kination domination eras.
The advantage of these inflationary models is that the spectral index takes values favorable by the Planck 2018 data. 
The PBH production from  runaway inflaton models has the unique and very attractive feature to automatically reheat the universe. In these models the PBHs are produced during the kination stage and their prompt evaporation efficiently produces the required entropy.
 Such runaway models are remarkably economic having  interesting implications for the early universe cosmology, possibly giving  rise to a wCDM late time cosmology as well.
 
\end{abstract}

\maketitle

\nopagebreak

\renewcommand{\thesection}{1}

\section{\label{introsection} Introduction}

The primordial origin of the black holes (PBH) is rather attractive scenario because PBHs %with sub-solar masses 
can constitute the entire 
cosmological dark matter (DM) or some significant fraction of it \cite{PBH1,PBH2,Carr:1975qj}.
Contrary to stellar black holes, the mass range of the PBHs can be very wide, spanning over thirty decades of mass, from $10^{-18} M_\odot$ to $10^{15} M_\odot$. 
%An exciting fact is that
 
 A synergy of observations, including the CMB  \cite{Ricotti:2007au}, the stochastic gravitational wave background \cite{Sasaki:2018dmp}, Lyman $\alpha$ forest \cite{Murgia:2019duy},  
  lensing events \cite{Barnacka:2012bm,Niikura:2017zjd,Tisserand:2006zx} and dynamical studies of bound astrophysical systems \cite{Capela:2013yf,Capela:2012jz,Brandt:2016aco,Graham:2015apa,Gaggero:2016dpq}
derive a combination of constraints on the PBH abundance for nearly the entire PBH mass range. 
Neglecting, possible enhanced merging rates or an extended PBH mass distribution the current allowed mass windows for the  dominant PBH dark matter scenario are quite narrow and have central values $M_\text{} \sim 10^{-15} M_\odot$ and $M_\text{} \sim 10^{-12} M_\odot$. 
Remarkably,  although triggered by the LIGO events \cite{TheLIGOScientific:2016pea, Abbott:2016blz, Abbott:2016nmj}, the PBH research has been extended into scenarios with vastly different mass scales.
PBHs with very light masses are anticipated to Hawking radiate energetically and this  places strong constraints on their abundance.  The lightest PBHs that can constitute a non-negligible part of the cosmic dark matter have mass $M\sim 10^{-17}M_\odot$. PBHs with smaller mass are prone to  evaporation and hence much constrained from the extra galactic gamma-ray background \cite{Carr:2016hva}, CMB and BBN \cite{Carr:2009}.
Mini PBHs with masses $M \ll 10^{-23}M_\odot \simeq 10^{10}$g  will have promptly evaporated in the very early universe potentially leaving no observational traces. 

However, the scenario of mini PBHs is of major interest due to the theoretical expectation that black holes cannot evaporate into nothing, see e.g. \cite{ Bowick:1988xh, Coleman:1991ku, Chen:2014jwq}. 
If the black hole evaporation halts at some point then a stable state, called black hole remnant, will survive.
Remnants from the PBHs prompt evaporation have important cosmological consequences, with the most notable one being  that PBH remnants can comprise the entire dark matter in the universe \cite{Barrow, Carr:1994ar}.  The cosmological scenario of mini PBHs evaporation and PBH remnants has been studied in several  contexts in \cite{Alexander:2007gj, Scardigli:2010gm, Lennon:2017tqq, Raidal:2018eoo, Rasanen:2018fom, Nakama:2018utx, Morrison:2018xla}.

%As aforementioned, 

The generation of mini PBHs implies that the ${\cal P_R}(k)$ has to feature a peak at very large wavenumbers $k$.
The most attractive mechanism to generate PBHs is inflation. 
Due to the natural generation of large scale perturbations from quantum fluctuations, inflation is the dominant paradigm that cosmologists follow to explain the origin of the large scale structure and has been, so far, successfully tested by the CMB precision measurements \cite{Akrami:2018odb}. Nevertheless inflation does not seed large scale perturbations only, it seeds perturbations in all scales. Hence, PBHs can form if perturbations strong enough to collapse are produced at scales $k^{-1} \ll k^{-1}_\text{cmb}$ characteristic of the PBH mass.

There have been  a numerous inflationary models  constructed to predict a significant abundance of PBHs, for recent proposal see e.g. 
\cite{Kawaguchi:2007fz, Drees:2011hb, Kawasaki:2016pql, Garcia-Bellido:2017mdw, Ballesteros:2017fsr, Hertzberg:2017dkh, Dalianis, Gao:2018pvq,Kawasaki:2016pql,Cicoli:2018asa, Ballesteros:2018wlw,Kannike:2017bxn, Gong:2017qlj, Pi:2017gih, Cai:2018tuh, Dimopoulos:2019wew}. Their common feature is that the spectrum of the curvature perturbations, ${\cal P_R}(k)$, turns from red into blue in small scales.
Since inflation models primarily generate the CMB anisotropies, these new inflationary proposals, though successful at generating PBHs, might fail at the $k^{-1}_\text{cmb}$ scale.
In general the predicted spectral index $n_s$ and running $\alpha_s$ values % predicted by the PBH generating inflationary  models 
are in best accordance with the CMB measured values when the PBHs have masses of minimal size. 
Light PBHs imply that the ${\cal P_R}(k)$ shape has to be modified at   $k$ values, far beyond the scales probed by the CMB. From a model building perspective, achieving an enhancement of the ${\cal P_R}(k)$ at the very end of the spectrum might seem an attractive  feature. This is an extra motivation to examine the mini PBH scenario.

In this work we build inflationary models that generate mini PBH and examine the early and late universe cosmology of the PBHs remnants. 
Our inflationary models belong in the family of the $\alpha$-attractors \cite{Kallosh:2013yoa}
and the building blocks we use are exponential potentials. 
The PBHs are generated by the presence of an inflection point at small field values where the inflaton velocity decreases significantly producing a spike in the  ${\cal P_R}(k)$.
Since, PBHs form during the very early post-inflationary cosmic stage and reheating might have not been completed,  it is natural to  examine the evolution of the PBHs and their remnants for different backgrounds and expansion rates. 

We derive expressions for the relic abundance of the PBH remnants for an arbitrary  barotropic parameter $w$ and remnant mass, $M_\text{rem}$. These expressions  are general and applicable to the stable PBH scenario as well. 
Afterwards, we examine explicitly  the radiation domination, the matter domination and the kination domination cases. We explicitly construct and analyze three different inflationary potentials and we compute  the power spectrum of the comoving curvature perturbation, solving numerically the Mukhanov-Sasaki equation, and we estimate the fractional PBH remnant abundance. A great advantage of our inflationary models is that the predicted values for the $n_s$ and $\alpha_s$ are placed inside the %sweet spot/ 
 1-$\sigma$ CL region of the Planck 2018 data. 

Moreover, we introduce the scenario of PBH production during the kination domination regime (called also stiff phase) that 
has interesting cosmological implications. 
Kination is driven by the kinetic energy of the  inflaton field itself. % that  also seeds the PBH formation.  
The duration of the kination regime is solely specified by  the mass and the abundance of the PBHs produced. 
The fact that the PBHs promptly evaporate means that the universe is automatically and successfully reheated 
 without the need of special couplings or tailor made resonance mechanisms.  
In addition, the inflaton field might play the role of the quintessence at late times giving rise to a testable wCDM cosmology, where the cold dark matter, comprised of PBHs remnants, is produced by the primordial fluctuations of the very same field. Apparently, in terms of ingredients, this is a maximally economic cosmological scenario.

The analysis in this paper is structured as follows. 
In Section 2  we discuss the bounds on the masses of the PBHs and of their remnants  reviewing briefly theoretical considerations and deriving the associated cosmological bounds.
In Section 3 the cosmology of the PBH remnants is presented for a general expansion rate and the main formulas are derived. 
In Section 4 we turn to the inflationary model building and formulate the constraints that the ${\cal P_R}(k)$ has to satisfy. In Section 5 we examine  inflationary models based on the $\alpha$-attractors that generate mini PBHs, we compute numerically the ${\cal P_R}(k)$, and construct explicit examples that the PBH remnants comprise the entire dark matter in the universe. We present and illustrate our results with several plots and tables.  In section 6 we conclude.

\begin{figure*}[htpb]
\centering
\includegraphics[width = 0.68\textwidth]{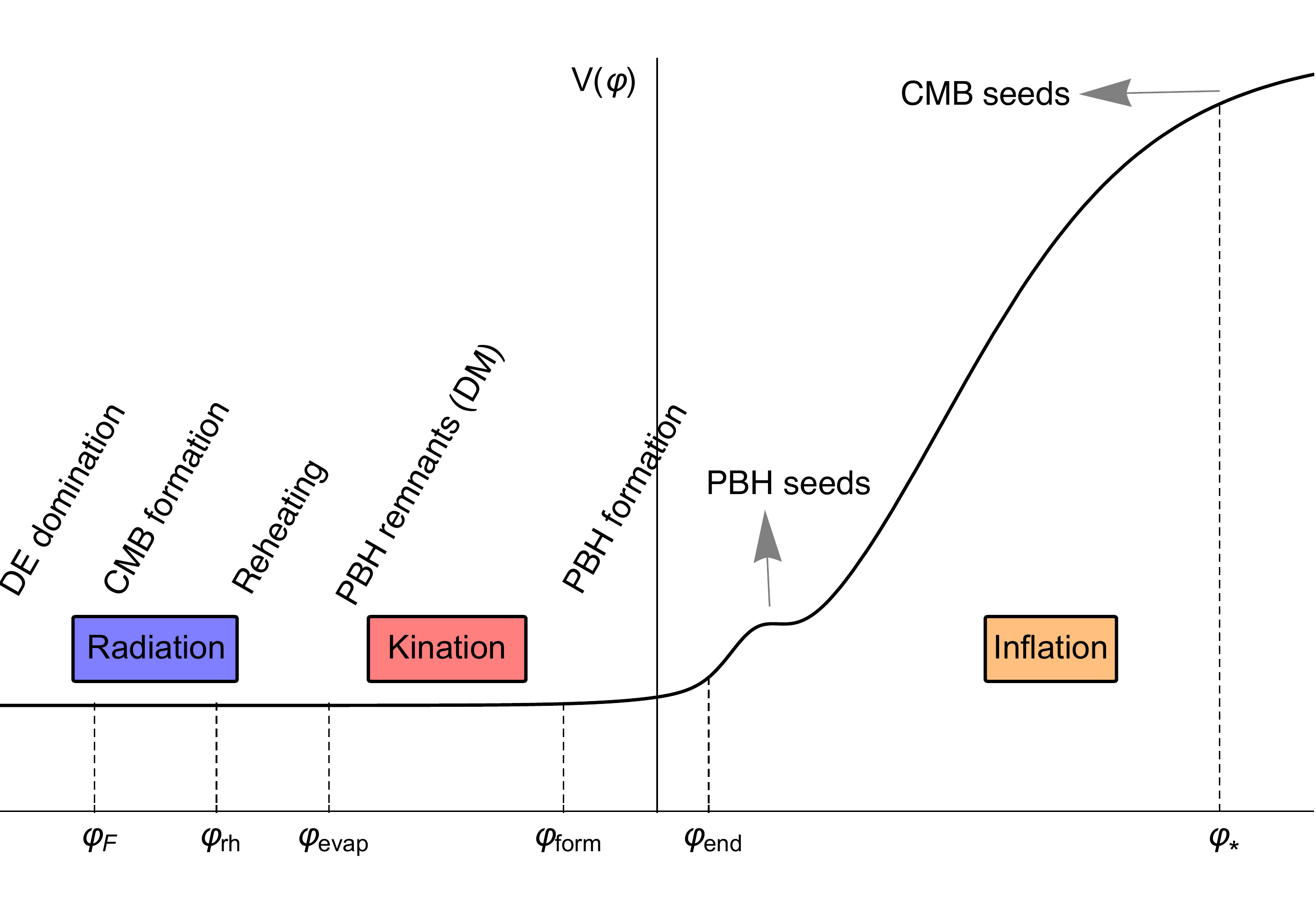}
\caption{
A schematic illustration of the runaway inflationary model introduced in this work (kination scenario) that produces PBHs, explains the dark matter with the PBH remnants,  reheats the Universe  via the PBH evaporation and implements a  wCDM late time cosmology.} \label{Fig1Vkin}
\end{figure*}

\renewcommand{\thesection}{2}
\section{PBH  evaporation remnants}

Hawking predicted that black holes radiate thermally with a temperature \cite{Hawking:1974sw, Hawking:1974rv}
\begin{align} \label{HawkT}
T_\text{BH}=\frac{\hbar c^3}{8\pi G M k_B} \sim  10^8 \left( \frac{M}{10^{5}\text{g}}\right)^{-1} \text{GeV}\,,
\end{align}
and are expected to evaporate on a time scale $t_\text{evap} \sim G^2M^3/(\hbar c^4)$, that is found to be \cite{Carr:2009}
\begin{align} \label{tevap}
t_\text{evap}= 407 \tilde{f}(M) \left( \frac{M}{10^{10} \text{g}} \right)^3 \text{s},
\end{align}
where $M$ the mass of the PBH formed.
We see that PBHs in the mass range $M\sim 10^{9}-10^{12}$ g evaporate during or after the BBN cosmic epoch and the $\beta$ is much constrained by the abundance of the BBN relics. For PBH in the mass range $M\sim 10^{13}-10^{14}$ g the evaporation takes place during the cosmic epoch of recombination and the CMB observations put the stringent constraints on the $\beta(M)$. Larger PBH masses contribute to the extra galactic gamma-ray background, for a review see  \cite{Carr:2009}.
 Hence, the scenario of the PBH remnants as dark matter is motivated for $M<10^9$ g. 
In particular, the remnants from the evaporation of the PBHs can constitute the entire or a significant portion of the cosmic dark matter if the PBH mass is smaller than 
 \begin{align}
 M <  \left( \frac{ \kappa \, m^2_\text{Pl} \, M^{1/2}_\text{eq}  }{1+w}\right)^{2/5}\,,
 \end{align}
 as we will show in the Section 3. The Planck mass is $m_\text{Pl}=2.2 \times 10^{-5}$ g and $M_\text{eq}$ is the horizon mass at the moment of radiation-matter equality. The $w$ stands for the equation of state of the background cosmic fluid. 
For reasons that will we explain later,  we call the upper mass bound $M_\text{inter}$ and it has size, roughly, $ 2 \, \kappa^{2/5} 10^6$ g. 
 Remnants from PBHs with mass $M>M_\text{inter}$ contribute only a small fraction to the total dark matter.

\subsection{Theoretical considerations about the PBH remnant mass}

There are several theoretical reasons for anticipating that black holes do not evaporate completely but leave behind a stable mass state.
The Hawking radiation is derived by treating matter fields quantum mechanically, while treating the space-time metric classically. When the mass of an evaporating black hole becomes comparable to the Planck scale  such a treatment would breakdown, and quantum gravitational effects would become relevant.
Energy conservation \cite{Torres:2013kda}, extra spatial dimensions \cite{ArkaniHamed:1998rs, Suranyi:2010yt},  higher order corrections to the action of general relativity \cite{Barrow}, the information loss paradox \cite{Chen:2014jwq}  could be sufficient to prevent complete evaporation.
Higher order correction to the Hawking radiation emerging from some quantum gravity theory are also expected to modify the evaporation rate. % the evaporation of black holes to stop. %  and allow BHs with mass $\lesssim 10^{-18}M_{\odot}$ leave a remnant  mass behind.
The  mass of the final state of the evaporation, i.e the PBH remnants, $M_\text{rem}$
can be written in terms of the Planck mass
%, is expected to be of the order of
\begin{align}
M_\text{rem}=\kappa \, m_\text{Pl}\,.
\end{align}
The $\kappa$ is a factor that parameterizes our ignorance. 
Different theories predict stable black hole relics of different mass. 
The $\kappa$ may be of order one, with relic black hole masses characterized by  the fundamental scale of gravity, $m_\text{Pl}=G^{-2}$,  but other values for the $\kappa$ are also admitted.
If black holes have quantum hair, e.g  if possess discrete electric and magnetic charges, the remnant mass depends on the value of the charge, $M_\text{rem}\sim m_\text{Pl}/g$, where $g$ the corresponding coupling constant \cite{Coleman:1991ku}. Thus, the $M_\text{rem}$ can be orders of magnitudes larger than one, $\kappa \gg 1$, for weakly coupled theories.
Other theories, such as those where a generalized uncertainty principle is applied \cite{Adler:2001vs} the mass of the black hole remnants can be much smaller that the $m_\text{Pl}$, see e.g \cite{Carr:2015nqa}, hence it is  $\kappa \ll 1$. 
In our analysis and expressions the $\kappa$ is a free parameter. 
This is a justified approach since we know nearly next to nothing about the physics at that  energy scales.
 In explicit examples  we will pick up  the benchmark  $\kappa \sim1$ value and we will comment on the cosmology of different $\kappa$ values, see Fig. \ref{betawk}.

This work aims at  the cosmology of the PBH remnants and we will remain agnostic about the fundamental physics that prevents black holes (or holes more properly) from complete evaporation. We will not enter into the details regarding the modification of the Hawking temperature with respect to the black hole mass either. 
Nevertheless, we remark that the formation of mini PBHs with mass $M<M_\text{inter} \sim \kappa^{2/5} 10^6$ g takes place in the very early universe, at the cosmic time $t_\text{form}$, and we expect these PBHs to evaporate promptly with their temperature reaching a maximum value contrary to what the standard expression (\ref{HawkT})
 dictates. 
If the temperature of the PBHs is initially smaller that the background cosmic temperature, $T_\text{BH}(t_\text{form})<T_\text{}(t_\text{form})$, the accretion effects should be taken into account. Although the accretion decreases the temperature of the PBH, the decrease of the cosmic temperature due to the expansion is much faster and the amount of matter that a PBH can accrete is small. Hence the PBH lifetime will not be  modified
and once the cosmic temperature falls below the value $T_\text{BH} \sim 10^8(M/10^5 \text{g})^{-1}$ GeV 
 %the accretion effects from the background radiation diminish and 
the PBHs heat up and evaporate. 
Concerning the black hole temperature, the $T_\text{BH}$ is expected to reach a maximal value and  afterwards decrease  as $M\rightarrow M_\text{rem}$. In this last stage of the PBH evaporation the rate $dM/dT_\text{BH}$ turns into positive.  %and the PBHs can exist in stable thermal equilibrium with the background radiation. %Hence the $M_\text{rem}$ might be time dependent determined by the  expansion dynamics. Nevertheless, 
The PBHs are expected to exist in stable equilibrium with the background only when their mass is already close to the remnant mass \cite{Barrow}, thus possible related corrections  can be considered negligible for the scope of this work.

\subsection{Cosmological constraints on the mass of the PBH remnants } 
 
The examination of the PBHs remnants  cosmology provides us with observational constraints on the $\kappa$  value. The corresponding analysis is presented in detail in the Section 3 and here, in advance,  we will use  part of the results to report the cosmological allowed values for the PBH remnants masses.

Let us first examine the minimal possible value for the $M_\text{rem}$.  
In the inflationary framework  
the formation of PBHs with mass $M$ can be realized only 
 if the horizon mass right after inflation, $M_\text{end}=4\pi M^2_\text{Pl}/H_\text{end} $ is smaller than $M/\gamma$. 
The $\gamma$ parameter is the fraction of the Hubble mass that finds itself inside the black hole. 
 In terms of the Hubble scale at the end of inflation, $H_\text{end}$, 
the bound reads
\begin{align} \label{Hcon}
{M}> \gamma\, 10^{5}\, \text{g} \,\frac{10^9 \,  \text{GeV}}{H_\text{end}}\, .
\end{align} 
The above inequality yields a lower bound for the PBH mass.
The upper bound on the tensor-to-scalar ratio $r_*<0.064$ \cite{Akrami:2018odb} and the measured value of the scalar power spectrum amplitude constrain the  $H_* \simeq (\pi^2 A_s r_*/2)^{1/2} M_\text{Pl}$. It is $H_\text{end}<H_*<2.6\times 10^{-5} M_\text{Pl}\simeq 6.5 \times 10^{13}$ GeV, hence the minimal PBH mass that can be generated is $M/\gamma > {\cal O}(1)  (r_*/0.06)^{-1/2}$ grams.
Assuming that a radiation domination phase follows inflation then the fractional abundance of the PBH remnants, given by Eq. (\ref{fRD2}), 
is maximal $f_\text{rem}=1$ for $\kappa \gtrsim 10^{-18.5}$.   
Hence, the PBH remnants are possible to have a significant relic abundance only if they have mass
\begin{align}
M_\text{rem} > 1 \, \text{GeV} \simeq 1.8\times 10^{-24}\text{g}\,.
\end{align}
This lower bound has been derived assuming the minimum possible PBH mass, $M\sim 1$ g, and the maximum possible formation rate, $\beta \sim 1$, see Eq. (\ref{fRD2}). It is also valid for non-thermal post-inflationary cosmic evolution.
 PBHs remnants with smaller mass can constitute only a negligible amount of the total dark matter energy density. 

On the other side, the $\kappa$ value has a maximum value, $\kappa_\text{max}=M/m_\text{Pl}$. Apparently for $\kappa =\kappa_\text{max}$ there is no Hawking radiation and one does not talk about PBH remnants. For $\kappa \ll M/m_\text{Pl}$
 the Hawking radiation is important and it might affect the BBN and CMB observables for light enough PBHs. Assuming again a radiation domination phase after inflation the  Eq. (\ref{fRD2}) implies that $\kappa \propto 1/ \beta$ for $f_\text{rem}=1$. Increasing the $\kappa$ smaller $\beta$ values 
 are needed. 
 According to the Eq. (\ref{tevap}), for $M \gtrsim 10^{9}$ g the PBH evaporate after the timescale of one second and the  BBN constrains $\beta<10^{-22}$, see Fig. \ref{FigbetaRD}. 
 However  the BBN $\beta$ upper bounds  cannot be satisfied for $\kappa \ll \kappa_\text{max}$. 
 Hence, for $M_\text{rem} \ll M$, we conclude that only  the remnants with mass in the window
 \begin{align}
10^{-24}\, \text{g} \, < M_\text{rem} \ll 10^{8} \text{g}
 \end{align}
can have a sufficient  abundance to explain the observed dark matter in the universe. The upper bound is determined by the BBN constraints on the parent PBH mass. It might be satisfied for $M_\text{rem}$ about one order of magnitude less than $10^8$ grams;  its exact value depends on how the  Eq. (\ref{HawkT}) and (\ref{tevap}) are modified and the equation of state $w$ after inflation.
 In the following we derive the expressions for the relic abundance of the PBH remnants for a general expansion rate and $M_\text{rem}$ parameter, and examine separately the cases of radiation, matter and kination  domination eras.

%For the inflationary model builders the essential is to design a model that features  a peak that generates PBHs in the preferred mass range $M<Minter$. 

%Moreover, the power spectrum at smaller $k$ has to be in accordance with the CMB and BBN constraints \cite{},

%The additional ingredient of the present analysis is that PBHs are not assumed to evaporate to nothing. 

%My constraints..if $k>>>1$ then bbn, cmb..

%It may also be that $\kappa \ll 1$

%\section{The position of the PS peak}

\renewcommand{\thesection}{3}
\section{The early universe cosmology of the PBH remnants}

The cosmology of the PBH remnants originating from large primordial inhomogeneities has been studied in detail in Ref. \cite{Barrow, Carr:1994ar} where the basic expressions have been derived.
PBH remnants might also originate form micro black holes produced from high energy collisions in the early universe \cite{Nakama:2018lwy}.
In the following we will consider the formation of  PBHs due to large inhomogeneities and generalize the key expressions for arbitrary barotropic parameter $w$, introducing also the PBH - stiff fluid (kination)  scenario. 

Let us suppose that at the early moment $t_\text{form}$
a fraction $\beta $ of the energy density of the universe collapses and forms primordial black holes.
The mass density of the PBHs is $\rho_\text{PBH}\simeq \gamma \beta \rho_\text{tot}$ at the moment of formation, where $\rho_\text{tot}=3H^2M^2_\text{Pl}$ with $M_\text{Pl}=m_\text{Pl}/\sqrt{8\pi}$, the reduced Planck mass. 
 %The $\gamma$ parameter is the fraction of the Hubble mass that finds itself inside the black hole.
The formation probability is usually rather small,  $\beta\ll 1$, and  the background energy density $\rho_\text{bck}=(1-\gamma\beta)\rho_\text{tot}$ is approximately equal to  $\rho_\text{tot}$.

The PBHs are pressureless non-relativistic matter and their number density $n_\text{PBH}$ scales like  $a^{-3}$. 
The background energy density scales as $\rho \propto a^{-3(1+w)}$ where $a(t)\propto t^{\frac{2}{3(1+w)}}$ and $w$ equation of state of the background fluid. The perturbations evolve  inside the curvature scale $1/H$, that has mass  $M_H=(3/4) (1+w) m^2_\text{Pl} t$, called the Hubble scale mass.
In the approximation of instantaneous evaporation, the moment right before evaporation, that we label $t^<_\text{evap}$, the energy density of the PBHs over the background energy density is 
\begin{align}
\frac{\rho_\text{PBH}(t^<_\text{evap})}{\rho_\text{bck}(t^<_\text{evap})}= \gamma \, \beta\, \gamma^\frac{2w}{1+w}\,\left( \frac{M_H(t^<_\text{evap})}{M}  \right)^\frac{2w}{1+w} \tilde{g}(g_*,  t_\text{evap})\,,
\end{align}
where $\tilde{g}(g_*, t_\text{evap})$ is equal to one unless 
the universe is radiation dominated; in that case it is $\tilde{g}(g_*,  t_\text{evap})\equiv (g_*(t_\text{form})/g_*(t_\text{evap}))^{-1/4}$ where $g_*$, the thermalized  degrees of freedom, that we took equal to the  entropic degrees of freedom, $g_s$.
Substituting the Hubble mass at the evaporation moment of a  PBH with mass $M$,
$M_H(t_\text{evap})=3 M^3(1+w)/4m^2_\text{Pl}$, a threshold  $\beta(M)$ value is found.  %(with a mild dependence on $\gamma$)
% that the universe becomes PBH dominated. 
For 
\begin{align} \label{betaMax}
\beta < \tilde{h}^{-1}(\gamma, w,  t_\text{evap}) \left(\frac{m_\text{Pl}}{M} \right)^\frac{4w}{1+w}\,,
\end{align}
where $\tilde{h}(\gamma, w,  t_\text{evap})\equiv \gamma^\frac{1+3w}{1+w}(\frac{3}{4}(1+w))^\frac{2w}{1+w} \tilde{g}(g_*,  t_\text{evap})$, the universe has never become PBH dominated. 

The moment right after the evaporation,  that we label $t^>_\text{evap}$, the energy density of the PBHs has decreased $(\kappa m_\text{Pl}/ M)^{-1}$ times. 
This factor is  much larger than one thus nearly the entire energy density of the initial PBHs turns into radiation apart from a tiny amount, reserved by the PBH remnants. 
The present density of the PBH remnants  
depends on the equation of state of the universe after the PBH evaporation. If we assume that a radiation domination phase follows the PBH evaporation 
the fractional abundance of the PBH remnants over the total  DM abundance  today is 
\begin{align} \label{f-reldom}
f_\text{rem}(M)=\tilde{c} (\gamma, w,  t_\text{eq})\left(\frac{M_\text{eq}}{M_H(t_\text{evap})} \right)^{1/2} \frac{\kappa\, m_\text{Pl}}{M}\, \left( \frac{M}{m_\text{Pl}}\right)^\frac{4w}{1+w}
\end{align}
where $\tilde{c}(\gamma, w, t_\text{eq})=2^{1/4}  \tilde{h}(\gamma, w, t_\text{eq})  {\Omega_\text{m}}/{\Omega_\text{DM}}$. 
However,  the assumption that there is a radiation domination phase after the PBH evaporation holds either when the universe has become PBH dominated at the moment $t^<_\text{evap}$ or when the equation of state of the background fluid is $w=1/3$.  Otherwise, one has to replace the $M_H(t_\text{evap})$ with the $M_\text{rh}$, that is the Hubble radius mass at the completion of reheating, and include a $w$-dependent factor to account for the different expansion rate. Next, particular cases will be examined.

\begin{figure}[htpb]
\centering
\includegraphics[width = 0.48\textwidth]{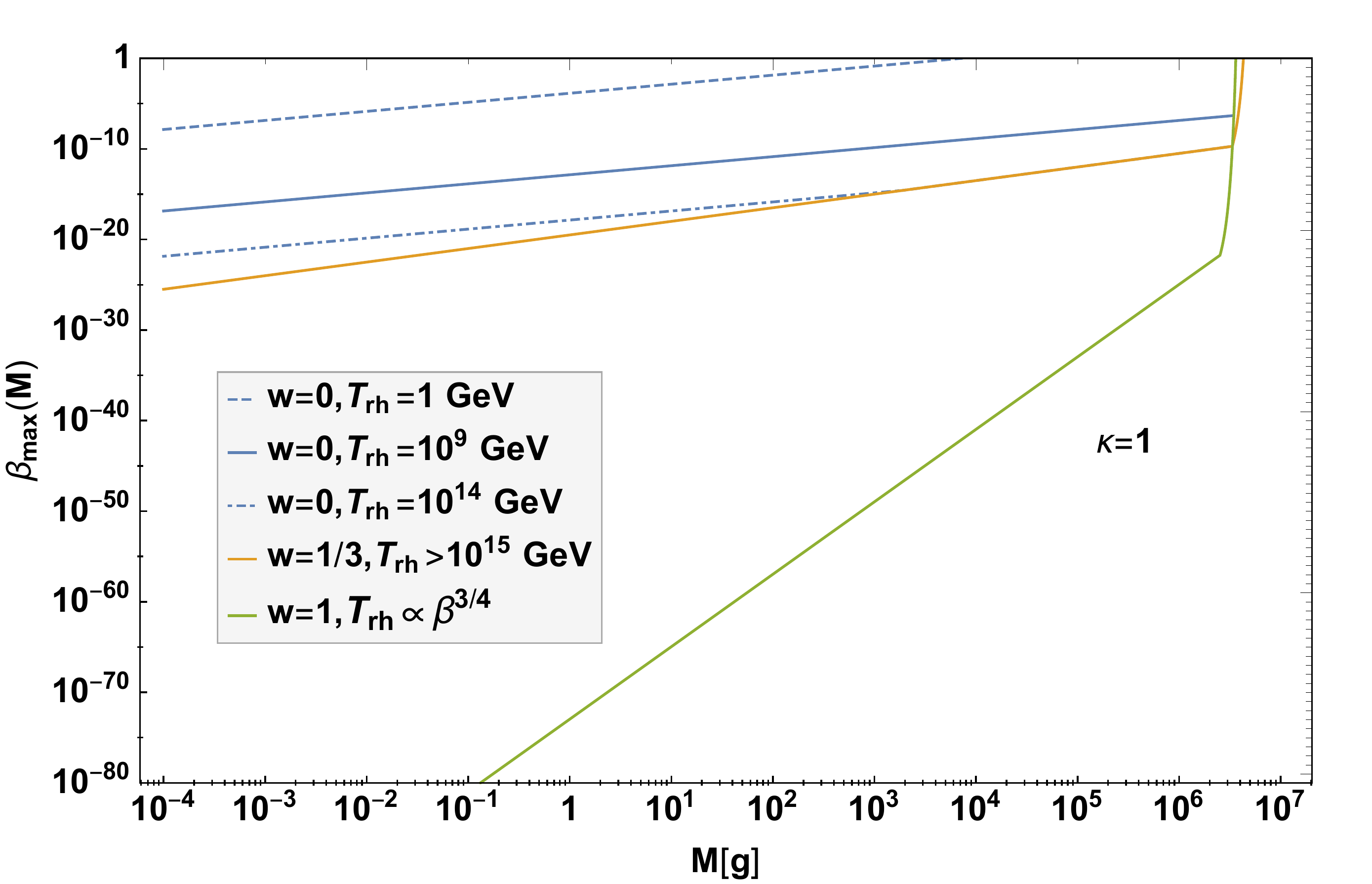}
\caption{The $\beta_\text{max}$, given by the condition $f_\text{rem}=1$ for PBH formation during three different cosmic phases with equation of states: $w=0$ (for three different reheating temperatures), $w=1/3$ and $w=1$. At the mass $M=M_\text{inter}$  the requirement $f_\text{rem}\leq 1$ does not imply any bound on the $\beta$.  We considered that the mass of the PBH relics is equal to the Planck mass, $\kappa=1$.} \label{betaw1}
\end{figure}

\begin{figure}[htpb]
\centering
\includegraphics[width = 0.48\textwidth]{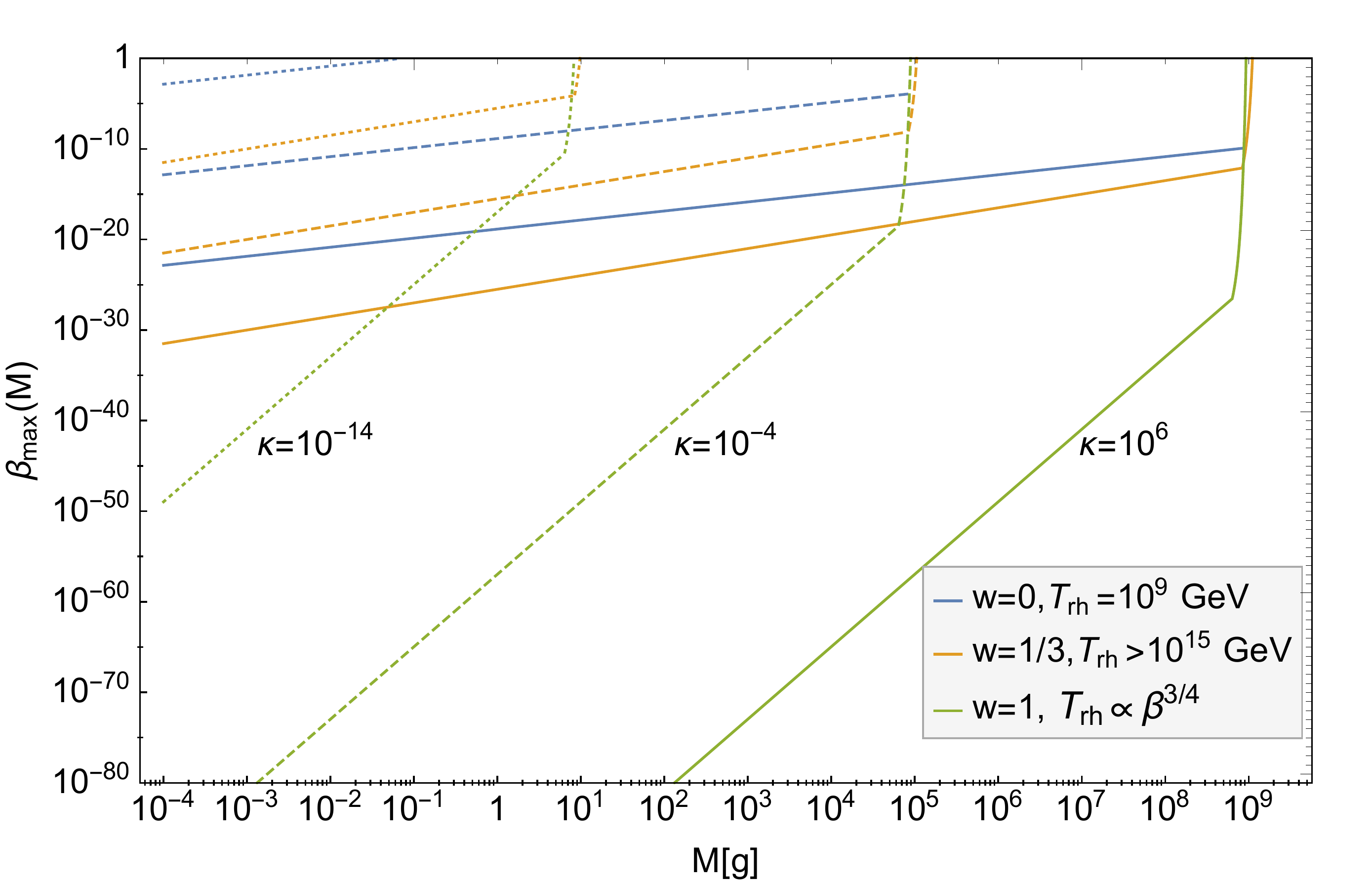}
\caption{s in Fig. \ref{betaw1}, for PBH relics with arbitrary chosen  masses $10^6 m_\text{Pl}$ (solid line), $10^{-4} m_\text{Pl}$ (dashed) and $10^2$ TeV (dotted). } \label{betawk}
\end{figure}

If the universe has become PBH dominated at the moment $t^<_\text{evap}$ 
and we ask for $f_\text{rem}(M)=1$ we get the mass
 \begin{align} \label{Minter}
 M_\text{inter}= \tilde{\alpha}^{2/5}(w) \left( \kappa \, m^2_\text{Pl} \, M^{1/2}_\text{eq} \right)^{2/5}\,,
 \end{align}
where $\tilde{\alpha}(w)=2^{1/4}  (\Omega_\text{m}/\Omega_\text{DM}) (\sqrt{3}(1+w))^{-1}$. 
This is the intersection mass of  Eq. (\ref{betaMax}), and the $f_\text{rem}(M)=1$ line, given by Eq. (\ref{f-reldom}). 
We see that the $M_\text{inter}$ slightly depends on the $w$. This means that 
there is a single PBH mass  that the
early universe becomes PBH dominated and the evaporation remnants account for the total DM,
%  and simultaneously explains the  $\Omega_\text{DM}$ value %without surpassing it 
for any positive value of the equation of state. % $w>0$.  
Plugging in values, $M_\text{eq}=6.9\times 10^{50}$g, $g_*(T_\text{eq})=3.36$ we obtain that $M_\text{inter} \simeq 2  \, \kappa^{2/5} 10^6$ g.
The intersection mass is the maximum $M$ value in the Fig. \ref{betaw1} and \ref{betawk}. For masses $M\geq M_\text{inter}$ the upper bound on $\beta$ is practically removed.
Turning to the $\beta$, the  $\beta_\text{inter}$ value that yields $f_\text{rem}=1$ 
and  momentarily PBH domination phase % at $t^<_\text{evap}$ 
is $w$-dependent,
\begin{align}
\beta_\text{inter}(w)=\tilde{h}^{-1}(\gamma, w)\,\tilde{\alpha}^{\frac{-2w}{5+5w}} \left(\frac{m_\text{Pl}}{M_\text{eq}} \right)^\frac{4w}{5+5w}\,.
\end{align}
Larger values for $w$ require smaller $\beta$, hence minimal $\beta$ values are achieved for $w=1$, see Fig. \ref{betaw1}, \ref{betawk}.

%For $M=M_\text{inter}$ and $\beta=\beta_\text{inter}$ % and in the approximation of instantaneous evaporation 
%the universe becomes momentarily, at $t=t^<_\text{evap}$, PBH dominated.
For masses $M>M_\text{inter}$ the relic abundance of the 
PBH remnants is always smaller than the total dark matter abundance even if  PBHs dominate the  early universe. 
% maximization of the PBH-relic abundance  requires a
%However this abundance 
Hence, for $M\geq M_\text{inter}$ there is no constraint on the $\beta$ from the $f_\text{rem}(M)$.
This can be understood as follows.  
Let us assume that $M=M_\text{inter}$ and $\beta=\beta_\text{inter}$ such that $\Omega_\text{rem}=\Omega_\text{DM}$. 
This means that
 right before evaporation % the universe is PBH dominated and 
the PBH number density is $n_\text{PBH}(t^<_\text{evap})\simeq \rho_\text{tot}/M_\text{inter}$. 
If it had been $\beta>\beta_\text{inter}$ the PBH-domination phase would have  started at times $t<t_\text{evap}$, but the number density of the PBH relics at the moment $t_\text{evap}$ would have been the same. 
Hence, the value of the $f_\text{rem}(M_\text{inter})$ does not increase for $\beta>\beta_\text{inter}$.
Also, for $M>M_\text{inter}$  the number density of the PBHs is always smaller than $\rho_\text{tot}/M_\text{inter}$  at the moment $t^<_\text{evap}\sim G^2 M^3_\text{max}$ even if $\beta(M)\sim 1$.
Since the PBHs will evaporate into PBH remnants with the universal mass 
$\kappa m_\text{Pl}$,
 the conclusion to be drawn is that the relic energy density parameter of  PBHs remnants with mass $M>M_\text{inter}$ is always less than $\Omega_\text{DM}$. 
Summing up, any constraint on the $\beta(M)$ for $M_\text{inter}<M<10^{17}$g comes only  from 
the Hawking radiation of the PBHs,  not from the abundance of the PBH remnants.

Let us now turn to the $M=M(k)$ relation assuming a one-to-one correspondence between the scale $k^{-1}$ and the mass $M$. 
Following the Press-Schechter  formalism \cite{Press:1973iz}, %at the moment $t_f$  the 
there is a probability $\beta$ an overdensity with wavenumber $k$ to collapse when it enters inside the Hubble radius (or some time later if the universe is matter dominated). 
The mass $M$  of the PBH  is related to the wavenumber $k=aH$
as
\begin{align} \label{ka1}
\frac{M}{M_\text{rh}} =\gamma \frac{H^{-1}}{H_\text{rh}^{-1}}
% =\gamma \left[\left(       \frac{k}{k_\text{rh}}       \right)^{\frac{-2}{3w+1}} 
%\right]^{\frac{3}{2}(1+w)} \\
=\gamma \left( \frac{k}{k_\text{rh}} \right)^{\frac{-3(1+w)}{3w+1}}\,,
\end{align}
 where we utilized the relation between the wavenumber and the scale factor,
 \begin{align} \label{ka}
\frac{k}{k_\text{rh}}=\left(\frac{a}{a_\text{rh}} \right)^{-\frac12(3w+1)}\,.
\end{align}
%\begin{align} \nonumber
%\frac{k}{k_\text{rh}}&= \frac{a}{a_\text{rh}}\frac{H}{H_\text{rh}} \\
%&=\left(\frac{a}{a_\text{rh}} \right)^{-\frac12(3w+1)}
%\end{align}
The horizon mass at the completion of reheating, $M_\text{rh}=4\pi \left({\pi^2 g_*}/{90}\right)^{-1/2} {M^3_\text{Pl}}/{T^2_\text{rh}}$, reads
\begin{align}
M_\text{rh} %=4\pi \left( \frac{\pi^2 g_*}{90}\right)^{-1/2} \frac{M^3_\text{Pl}}{T^2_\text{rh}}\,
\simeq 10^{12} \, \text{g}\, \left( \frac{T_\text{rh}}{10^{10}\, \text{GeV}}\right)^{-2} \left(\frac{g_*}{106.75} \right)^{-1/2}\,.
\end{align}
If PBHs form during radiation domination era it is $M/\gamma>M_\text{rh}$ whereas if they form before the completion of the thermalization of the universe it is $M/\gamma<M_\text{rh}$. 
%Let us call $\tilde{N}_\text{rh}$ the e-folds of the reheating stage 
%then the wavenumber at the moment of reheating completion is $k_\text{rh}=k_\text{end}e^{-\frac{3w+1}{2}\tilde{N}_\text{rh}}$. The $k_\text{end}$ is the wavenumber that inflation ends and it is related to the Planck pivot scale $k_*=0.05$ Mpc$^{-1}$ as $k_\text{end}=k_*e^{N_*} (H_\text{end}/H_*)$. The $N_*$ are the e-folds of the observable inflation,
%\begin{equation} \label{Nrh}
%N_* \simeq 57.6 +\frac14 \ln \epsilon_* +\frac14 \ln \frac{V_*}{\rho_\text{end}} -\frac{1-3w_\text{}}{4} \tilde{N}_\text{rh} \,
%\end{equation}
%where $\epsilon_*$, $H_*$, $V_*$ are respectively the first slow-roll parameter, the Hubble scale and the potential energy when the CMB pivot scale exits the Hubble radius, while $H_\text{end}$, $\rho_\text{end}$ are the Hubble scale and the energy density at the end of inflation. 
We will return to the relation between the PBH mass and the wavenumber  $k$ in the Section 4, where we will explicitly write the $M=M(k, T_\text{rh}, w)$ formula in order to connect the PBH mass with the ${\cal P_R}(k)$ peak.

Let us note the formation of PBHs with mass $M$ is possible only if
the horizon mass right after inflation  is smaller than $M/\gamma$, see Eq. (\ref{Hcon}). 
Equivalently, a  PBH with mass $M$ will form due to superhorizon perturbations only if the  corresponding wavelength $k^{-1}$ is larger than the Hubble scale at the end of inflation. Thus, a different way to express the condition (\ref{Hcon}) is $k_\text{end}>k$.

Next, we examine separately the interesting cosmological scenarios with barotropic parameter $w=0$, $w=1/3$ and $w=1$.

\subsection{PBH production during radiation domination}

Let us assume that the bulk energy density is in the form of radiation.
Thus,  it is $\rho_\text{PBH}\simeq \gamma_\text{} \beta \rho_\text{rad}$ after the approximation $\rho_\text{rad}=(1-\beta)\rho_\text{tot} \simeq \rho_\text{tot}$, that is legitimate for $\beta \ll 1$. 
The PBH mass is $M=\gamma_\text{} M_H$  where 
 $M_H=m^2_\text{Pl}/(2H) \simeq m^2_\text{Pl} t$ is the Hubble radius mass during radiation domination (RD).
 
 Assuming a RD phase until the moment of the evaporation and making the approximation of instantaneous evaporation,  the energy density of the PBHs at the moment  right before evaporation is $\rho_\text{PBH}(t^<_\text{evap})=\gamma^{3/2}_\text{}\beta M\,m^{-1}_\text{Pl} \rho_\text{rad}(t^<_\text{evap})$. Thus, the assumption of a radiation dominated phase is valid for $\gamma^{3/2}_\text{} \beta M \,m^{-1}_\text{Pl} <1$.
In the opposite case the universe becomes PBH dominated before the moment of evaporation.
At the moment right after the PBH evaporation the energy density of the PBH relics is $\kappa  \gamma^{3/2}_\text{} \beta$ times the energy density of the radiation background.
For an RD phase until the epoch of matter-radiation equality, $t_\text{eq}$, it is  
$\rho_\text{rem}(t_\text{eq}) \simeq (\kappa m_\text{Pl}/M)\gamma \beta  \rho_\text{rad} T_\text{evap}/T_\text{eq}$ and the fractional abundance of the PBH remnants is found, 
\begin{align} \label{fRD}
f_\text{rem}(M) =\tilde{c}_\text{R}\,\gamma^{3/2} \,\frac{\kappa\, m_\text{Pl}}{M}   \beta \, \left(\frac{M_\text{eq}}{M_\text{}}\right)^{1/2} 
\end{align}
where  $\tilde{c}_\text{R}=2^{1/4}  \left({g(T_f)}/{g(T_\text{eq})} \right)^{-1/4}{\Omega_\text{m}}/{\Omega_\text{DM}}$. 
The $T_\text{evap}$ and $T_\text{eq}$ are the cosmic temperatures at the moment of evaporation and the epoch of matter-radiation equality.
The effectively massless degrees of freedom for the energy and  entropy densities were taken to be equal.
The maximum value $f_\text{rem}=1$, gives the maximum value for the $\beta_\text{max}(M)$, see Fig. \ref{betaw1} and \ref{betawk}.
The Eq. (\ref{fRD}) rewrites after inserting benchmark values,
\begin{equation} \label{fRD2}
f_\text{rem} (M)\, \simeq\,\kappa \left(\frac{\beta_\text{}}{10^{-12}}\right) \, 
 \Big(\frac{\gamma_\text{}}{0.2}\Big)^{\frac{3}{2}} 
\left(\frac{M}{10^{5}\text{g}}\right)^{-3/2}\,
\end{equation}
where we omitted the factor $0.95  \left({g(T_{k})}/{106.75}\right)^{-\frac{1}{4}} $ from the r.h.s. and took $\Omega_{\text{DM}}h^2=0.12$.

Assuming Gaussian statistics, the black hole formation probability for a spherically symmetric region is 
\begin{equation} \label{brad}
\beta_\text{}(M)=\int_{\delta_c}d\delta\frac{1}{\sqrt{2\pi\sigma^2(M)}} e^{-\frac{\delta^2}{2\sigma^2(M)}}\,
\end{equation}
that is approximately equal to  $\beta \sqrt{2\pi} \simeq \sigma(M)/\delta_c e^{-\frac{\delta^2_c}{2\sigma^2(M)}} $. The PBH abundance has an exponential sensitivity to the variance of the perturbations $\sigma(k)$ and to the threshold value $\delta_c$.   
In the comoving gauge Ref. \cite{Harada:2013epa} finds that the $\delta_c$ has the following dependence on the $w$,
\begin{equation} \label{dc}
\delta_c = \frac{3(1+w)}{5+3w}\sin^2 \frac{\pi \, \sqrt{w}}{1+3w}\,.
\end{equation}
For $w=1/3$ it is $\delta_c=0.41$. 
The variance of the density perturbations in a window of $k$ is given by the relation $\sigma^2 \sim {\cal P}_{\delta}$ where the ${\cal P}_{\delta}$ is related to the power spectrum of the comoving curvature perturbation as 
\begin{align}
  {\cal P_R}=\left( \frac{5+3w}{2(1+w)}\right)^2 {\cal P}_{\delta}\,,
\end{align}
hence, it is $\sigma^2 \sim (4/9)^2   {\cal P_R}$.
From the approximation of the Eq. (\ref{brad})  we get that ${\cal P_R} \sim (9/4)^2 (\delta^2_c/2) \ln(1/(\sqrt{2 \pi} \beta))^{-1}$. Benchmark values $\beta=10^{-12}$, $\delta_c=0.41$, $\kappa=1$ yield the required value for the power spectrum ${\cal P_R} \sim 1.6\times 10^{-2}$. Increasing the value of $\kappa$ by one  orders of magnitude or more gives only a slight decrease in the required value of the ${\cal P_R}$. 

Finally, let us note that  PBHs are expected to form with mass $M=\gamma_\text{} M_H$ when the cosmic temperature is
\begin{align} \label{TM}
T(M) \simeq 10^{11} \, \text{GeV}\, \gamma^{1/2}_\text{}\, \left(\frac{M}{10^{10}\,\text{g}}\right)^{-1/2} 
\left( \frac{g_*}{106.75}\right)^{-1/4}\,\,.
 \end{align}   
For example, formation of PBHs with mass $M\sim 10^5$ g requires reheating temperatures $T_\text{rh}>10^{13}$ GeV.
If the reheating temperature is lower than  $T(M)$  then PBHs with mass $M$ form during a non-thermal phase.

A particular example that yields $f_\text{rem}=1$ is described in the Section 5 and the $\beta(M)$ 
is depicted in Fig. {\ref{FigbetaRD}}.

\subsection{PBH production during matter domination}

For PBH formation during matter domination era (MD) the expression (\ref{fRD})
has to be multiplied with $(t_\text{form}/t_\text{rh})^{1/2}$ to account for the absence of a relative redshift of the $\rho_\text{PBH}$ with respect to the background energy density. It is,
\begin{align} \label{fMD}
f_\text{rem}(M, M_\text{rh}) =\tilde{c}_\text{M}\,\frac{\kappa\, m_\text{Pl}}{M} \gamma_\text{} \,  \beta \, \left(\frac{M_\text{eq}}{M_\text{rh}}\right)^{1/2}  \,,
%\nonumber \\
%(\simeq 10^{19}\, \frac{\kappa\, m_\text{Pl}}{M} \gamma_\text{} \,  %\beta \, \frac{T_\text{rh}}{10^{10}\text{GeV}})\,,
 \end{align}
where $\tilde{c}_\text{M}=2^{1/4} \left({g(T_\text{rh})}/{g(T_\text{eq})} \right)^{-1/4} \Omega_\text{m}/\Omega_\text{DM}$
and for $M<M_\text{rh}$. 
The Hubble scale mass at the completion of reheating reads $M_\text{rh}=M_H (T_\text{rh}, g_*)=4\pi \left( {\pi^2 g_*}/{90}\right)^{-1/2} {M^3_\text{Pl}}/{T^2_\text{rh}} $. 
The Eq. (\ref{fMD}) rewrites after normalizing the $M$, $M_\text{rh}$ with benchmark values,
\begin{align}
f_\text{rem}(M, M_\text{rh}) \nonumber
\simeq  3 \, \kappa \,& \gamma_\text{} \,  \left(\frac{\beta}{10^{-9}}\right) \, 
\left(\frac{M_\text{rh}}{10^{10}\text{g}} \right)^{-1/2}  \\
&\times \left( \frac{M}{10^5 \text{g}} \right)^{-1 } \left( \frac{g_*}{106.75}  \right)^{-1/4}
\end{align}
and the mass $M$ is related to the scale $k^{-1}$ of the inhomogeneity as
\begin{align} \label{kmd1}
k_\text{}=k_\text{end}\,  \left( \frac{4\pi M^2_\text{Pl}}{H_\text{end}}\right)^{1/3} \left(\frac{M}{\gamma_\text{}}\right)^{-1/3}\,, \quad \text{for} \,\quad
k_\text{}>k_\text{rh}\,.
\end{align}
In a pressureless background overdensities can easier grow and collapse if the MD era is sufficiently long. Contrary to the RD case non-sphericity and spin effects suppress the formation probability. 
Ref. \cite{Harada:2016mhb} examined the PBH production in MD era and found that for not very small $\sigma$ the PBH production rate  tends to be proportional to $\sigma^5$, 
\begin{equation} \label{bmat}
\beta_\text{}(M)\, = \, 0.056 \, \sigma^5\,,
\end{equation} 
If the collapsing region has  angular momentum the formation rate is further suppressed and reads \cite{Harada:2017fjm},
\begin{equation} \label{bspin}
\beta_\text{}(M)=2\times 10^{-6}f_q(q_c) {\cal I}^6 \sigma^2 e^{-0.147\frac{ \, {\cal I}^{4/3}}{ \sigma^{2/3}}}\,.
\end{equation}
Benchmark values are $q_c=\sqrt{2}$, ${\cal I}=1$, $f_q \sim 1$. According to \cite{Harada:2017fjm} the expression (\ref{bspin}) applies for 
 $\sigma \lesssim 0.005$  whereas Eq.  (\ref{bmat}) applies for $0.005 \lesssim \sigma \lesssim 0.2$.

During MD era, an additional  critical parameter is the duration of the gravitational collapse. Ref.  \cite{Harada:2017fjm} concluded that
the finite duration of the PBH formation can be neglected if the reheating time $t_\text{rh}$ satisfies $t_\text{rh} >\left( \frac{2}{5} {\cal I} \, \sigma\right)^{-1} t_k$
 where $t_k$ is the time of the horizon entry of the scale $k^{-1}$ (it does not coincide with the formation time $t_\text{form}$). 
 In terms of  temperatures this condition rewrites $T_\text{rh}< \left( \frac{2}{5} {\cal I} \, \sigma\right)^{1/2} T_k$
where $T_k$ the temperature that the scale $k^{-1}$ would enter the horizon during RD. 
Let us define the  temperature 
\begin{align}
T^{\text{MD}}_\text{form}= \left( \frac{2}{5} {\cal I} \, \sigma\right)^{1/2} T_k\,.
\end{align}
If the reheating temperature is smaller than $T^{\text{MD}}_\text{form}$ then PBHs form during MD era.
Unless this conditions is fulfilled the time duration for the overdensity to grow and enter the nonlinear regime is not adequate.
%Due to the fact that the collapse does not happen instantaneously after the horizon crossing  
Hence, the formation rates (\ref{bmat}), (\ref{bspin}) apply only for the scales $k$
 that experience a variance of the  comoving density contrast at horizon entry that is larger than 
 \begin{align} \label{sigmaCr}
 \sigma > \sigma_\text{cr} \equiv \frac52 {\cal I}^{-1} \left( \frac{k_\text{rh}}{k}\right)^3\,.
 \end{align}
In terms of temperature this translates into $\sigma>5/2 \, {\cal I}^{-1} (T_\text{rh}/T)^2$. If $\sigma<\sigma_\text{cr}$ one should consider the radiation era formation rate. % and in our numerics we will choose ${\cal I}=1$. (*Later *)

A particular example that yields $f_\text{rem}=1$ is described in the Section 5 and the $\beta(M)$ 
is depicted in Fig. \ref{FigbetaMspin}.
The $\sigma$ for that example is less than $0.005$ and larger than $\sigma_\text{cr}$, thus the overdensity collapses during the matter era with the  spin effects being crucial.
% We comment that the observational constraints, for $M>M_\text{rh}$, e.g. the BBN or the CMB, are $(M/M_\text{rh})^{1/2}$ less stringent than in the radiation case.

\subsection{PBH production during stiff fluid domination}

Let us  assume that the bulk energy density is in the form of stiff fluid (SD era), that is a fluid with barotropic parameter $w=1$, also called kination phase.
A non-oscillatory inflaton can give rise to a kination phase. 
It is $\rho_\text{PBH}\simeq \gamma_\text{} \beta \rho_\text{S}$ where we approximated $\rho_\text{S}=(1-\beta)\rho_\text{tot} \simeq \rho_\text{tot}$ for $\beta \ll 1$. 
The PBH mass is $M=\gamma M_H$  where 
 $M_H=(3/2)M^3/m^2_\text{Pl}$ is the Hubble scale mass for stiff fluid domination. 

 Assuming that the SD era  lasts at least until the moment of the evaporation and making the approximation of instantaneous evaporation,  the energy density of the PBHs at the moment right before evaporation is 
\begin{align} \label{KinEvap}
\frac{ \rho_\text{PBH}(t^<_\text{evap})}{\rho_\text{S}(t^<_\text{evap})}=\frac{3}{2}\gamma^{2}_\text{} \beta \frac{M^2}{ m^{2}_\text{Pl}} \,.
\end{align}  
The assumption of a kination phase is valid roughly for $\gamma^{2}_\text{} \beta M^2 \,m^{-2}_\text{Pl} <1$, otherwise the universe becomes PBH dominated before the moment of evaporation.

Let us assume that $\gamma^{2}_\text{} \beta M^2 \,m^{-2}_\text{Pl} <1$. 
At the moment right after the PBH evaporation the energy density of the PBH remnants is $\rho_\text{rem}(t^>_\text{evap}) =(3/2)\kappa  \gamma^{2}_\text{} \beta (M/m_\text{Pl}) \rho_\text{tot}$. The background energy density is now partitioned between the stiff fluid, $\rho_\text{S}$, and the entropy produced by the PBH evaporation, $\rho_\text{rad}$. The later is about $M/(\kappa m_\text{Pl})$ times larger than the $\rho_\text{rem}(t^>_\text{evap})$. Assuming that the evaporation products thermalize fast, the radiation redshifts like $\rho_\text{rad} \propto g_* g_s^{-4/3} a^{-4}$ whereas the stiff fluid background redshifts like $\rho_\text{S} \propto a^{-6}$. 
At  some  moment the radiation dominates the background energy density and we define it as the {\it reheating} moment $t_\text{rh}$. The scale factor is
\begin{align}
\frac{a(t_\text{rh})}{a(t_\text{evap})} =\left( \frac{2}{3}\frac{m^2_\text{Pl}}{M^2}  \frac{1}{\gamma^2_\text{} \beta}\,  \frac{g_*^{1/3}(t_\text{rh})}{g_*^{1/3}(t_\text{evap})}
\right)^{1/2}\,.
\end{align}
At that moment we also define the reheating temperature of the universe that reads 
\begin{align} \label{Tkination}
T_\text{rh} \equiv 6.3\, \text{MeV} \left( \frac{\beta}{10^{-28}} \right)^{3/4} \gamma^{3/2}_\text{}  g_*^{-1/2} \,.
\end{align}
Until the moment  $t_\text{rh}$ the energy density of the PBH remnants increases relatively to the stiff fluid dominated background as $\rho_\text{rem}/\rho_\text{S}\propto a^3$ and afterwards, that radiation dominates, it increases as $\rho_\text{rem}/\rho_\text{rad}\propto T^{-1}$.  
It is
\begin{align} \label{frelS1}
f_\text{rem}(M)=  \tilde{c}_\text{S} \frac32 \,  \gamma^{2}_\text{} \, \beta \, \frac{\kappa \,M}{m_\text{Pl}}\left(\frac{a(t_\text{rh})}{a(t_\text{evap})} \right)^3 \left(\frac{M_\text{eq}}{M_\text{rh}}\right)^{1/2}
\end{align}
where $\tilde{c}_\text{S}=2^{1/4} \left({g(T_\text{rh})}/{g(T_\text{eq})} \right)^{-1/4} \Omega_\text{m}/\Omega_\text{DM}$
and $M_\text{rh}=M_H(t_\text{rh})$.
For times $t<t_\text{rh}$ the Hubble radius mass increases like $M_H \propto a^3$ and, given that $M_H(t_\text{evap})=(3/2)M^3/m^2_\text{Pl}$, we find the $M_\text{rh}$ mass
\begin{align}
M_\text{rh}=\sqrt{\frac{2}{3}}\,  \gamma^{-3}_\text{} \, \beta^{-3/2} \, g^{1/2}_* \, m_\text{Pl}\,.
\end{align}
 Therefore, the Eq. (\ref{frelS1})  rewrites
\begin{align} \label{frelS2}
f_\text{rem}(M)=  \sqrt{\frac23}\,\tilde{c}_\text{} \left(\frac32 \gamma^{2}_\text{} \, \beta\right)^{1/4} \kappa  \left(\frac{m_\text{Pl}}{M} \right)^{3/2}
 \left(\frac{M_\text{eq}}{M}\right)^{1/2}
\end{align}
and normalizing with benchmark values we attain
\begin{align} \label{fremKIN}
f_\text{rem}(M) 
\simeq 4  \, \kappa \, \sqrt{\gamma_\text{}} \,  \left(\frac{\beta}{10^{-32}}\right)^{1/4} \, 
\left( \frac{M}{10^5 \text{g}} \right)^{-2 }.
\end{align}
For $\kappa \sim 1$ and $M\sim 10^5$g,  $\beta$ values as small as $10^{-32}$ can explain the observed dark matter in the universe. Pressure is maximal and we expect the overdense regions to be spherically symmetric. Utilizing the relation (\ref{dc}) the PBH formation occurs when the density perturbation becomes larger than $\delta_c=0.375$ and for the formation probability $\beta(M)$  given by the Eq. (\ref{brad}) we find that power spectrum values ${\cal P_R} \lesssim3.5 \times 10^{-3}$, for $\kappa \gtrsim 1$ 
and $M\sim 10^{5}$g, can yield $f_\text{rem}=1$.

\begin{figure}[htpb]
\centering
\includegraphics[width = 0.48\textwidth]{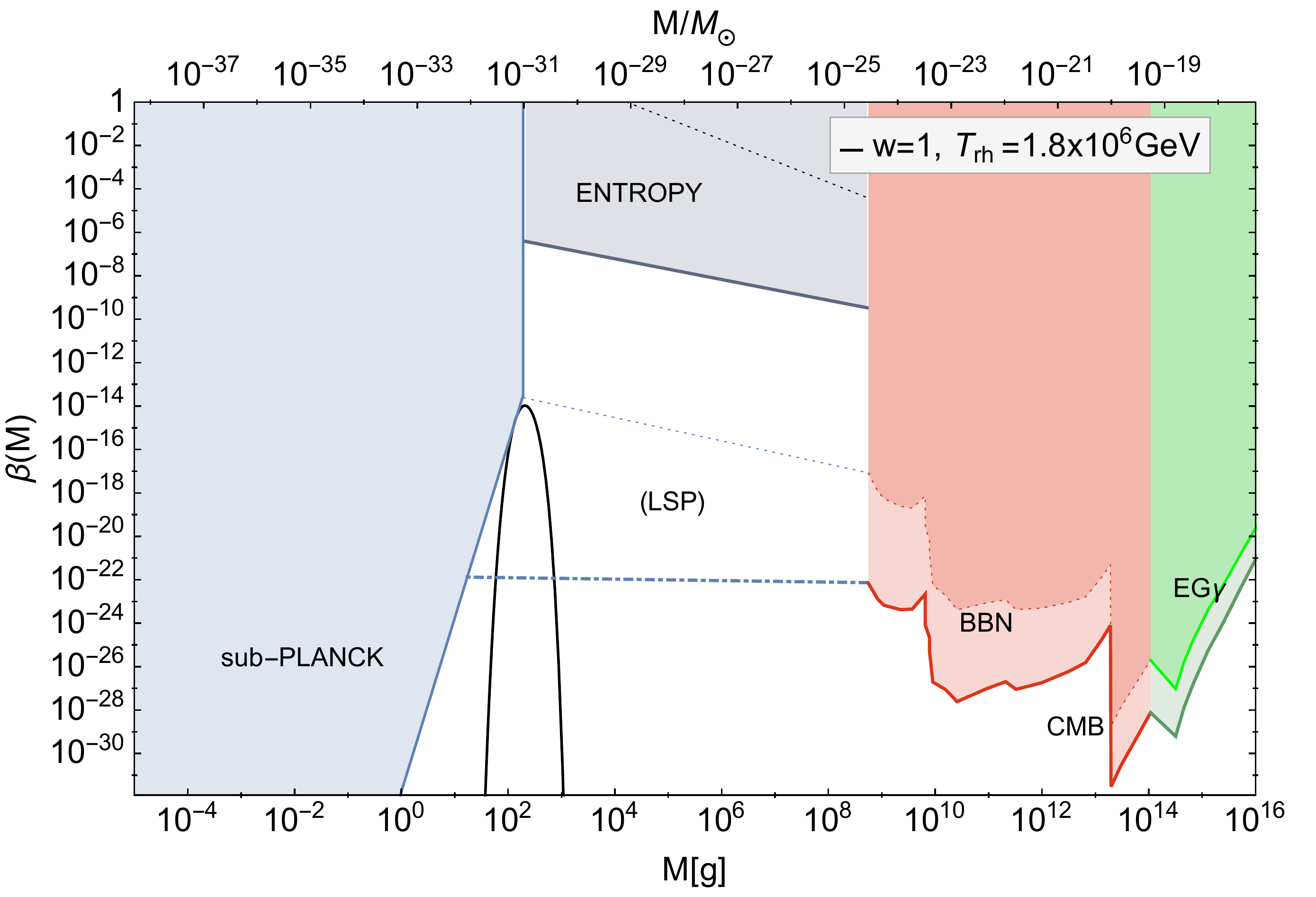}
\caption{
The $\beta(M)$ predicted by the inflationary model  (\ref{Vrun}) with parameters listed in the Table III for the  SD case. 
The PBHs with central mass $M= 2\times 10^{2}$ g are produced and evaporate during the kination era (SD). The evaporation leaves behind subplanckian mass remnants with mass $M_\text{rem}= 4\times 10^{-11}  m_\text{Pl} \simeq 6\times 10^8$ GeV  that comprise the entire dark matter, $f_\text{rem}=1$.
The reheating temperature is determined by the Eq. (\ref{Tkination}) and is proportional to $\beta^{3/4}$.
The  $\beta_\text{max}$ constraints for the entropy, BBN, CMB and EG$\gamma$, are  $(M/M_\text{rh})^{1/2}$  
stringent compared to the RD case (dotted lines).
} \label{betaw2}
\end{figure}

A particular example that yields $f_\text{rem}=1$ is described in the Section 5 and the $\beta(M)$ 
is depicted in Fig. \ref{betaw2}. 
%Note that the observational constraints, for $M>M_\text{rh}$, e.g. the BBN or the CMB, are $(M/M_\text{rh})^{1/2}$ more stringent than in the radiation case.

\subsubsection*{BBN constraints}

A kination regime has to comply with the  BBN constraints. Let us assume that a runaway inflaton $\varphi$ is responsible for the kination regime.  The energy density during BBN is partitioned between the kinetic energy of the $\varphi$ field and the background radiation.  Any modification to the simple  radiation domination regime
is parameterized by an equivalent number of additional neutrinos and the Hubble parameter has to satisfy the constraint  \cite{Simha:2008zj}
\begin{align}
\left. \left( \frac{H}{H_\text{rad}}\right)^2 \right|_{T=T_\text{BBN}} \leq 1+\frac{7}{43}\Delta N_{\nu_\text{eff}}\simeq 1.038
\end{align}
where $H$ the actual Hubble parameter and $H_\text{rad}$ the Hubble parameter if the total energy density was equal to the radiation. The $\Delta N_{\nu_\text{eff}}=3.28-3.046$ is the difference between the cosmologically measured value and the SM prediction for the effective number of neutrinos.  
In order to prevent the universe from expanding too fast during BBN  due the extra energy density $\dot{\varphi}^2/2$ % of the  kination regime occurs in the early universe,
 the reheating temperature has to be larger than \cite{Artymowski:2017pua} 
\begin{align} \nonumber
T_\text{rh}>&(\alpha-1)^{-1/2} \left( \frac{g_*(T_\text{BBN})}{T_\text{rh}}\right)^{1/4} T_\text{BBN} \\
&= {\cal O}(10)\, \text{MeV}
\end{align} 
where $\alpha \equiv 1+7/43 \Delta N_{\nu_\text{eff}} \simeq 1.038$.

An additional issue is that the gravitational wave energy in the GHz region gets enhanced during the kination regime \cite{Giovannini:1999bh, Riazuelo:2000fc, Yahiro:2001uh, Boyle:2007zx}.  
The energy density of the gravitational waves does not alter BBN predictions if
\begin{align}
I \equiv h^2 \int^{k_\text{end}}_{k_\text{BBN}} \Omega_\text{GW}(k)d \ln k \leq 10^{-5}
\end{align} 
which is written as \cite{Dimopoulos:2017zvq}
\begin{align}
I=\frac{2\epsilon h^2 \Omega_\text{rad}(t_0)}{\pi^{2/3}}\left(\frac{30}{g(T_\text{reh})} \right)^{1/3}\frac{h^2_\text{GW} V^{1/3}_\text{end}}{T^{4/3}_\text{rh}}
\end{align}
where $\epsilon\sim 81/(16\pi^3)$, $ h^2 \Omega_\text{rad}(t_0)=2.6 \times 10^{-5}$ and $h^2_\text{GW}=H^2_\text{end}/(8\pi M^2_\text{Pl})$.
Substituting numbers, the observational constraint $I \lesssim 10^{-5}$ gives a lower bound on the reheating temperature,
\begin{align} \label{TGW}
T_\text{rh} \gtrsim 10^{6}\, \text{GeV} \left( \frac{106.75}{g_*}\right)^{1/4} 
\left(\frac{H^2_\text{end}}{10^{-6}M_\text{Pl}} \right)^2\,.
\end{align}
Substituting the reheating temperature predicted by kination-PBH models, Eq. (\ref{Tkination}), into the bound (\ref{TGW}) we obtain a lower bound on the formation rate  
\begin{align}
\beta \gtrsim 2\times 10^{-17}  \left( \frac{106.75}{g_*}\right)^{1/3} \left(\frac{H_\text{end}}{10^{-6} M_\text{Pl}} \right)^{8/3}\,.
\end{align}
Smaller values for the $\beta$ mean that the radiation produced from the PBH evaporation dominates later during the early cosmic evolution and the kination regime is dangerously extended.  Asking for $f_\text{rem}=1$ the lower bound on the $\beta$ yields a lower bound on the  ratio 
\begin{align}
\frac{M}{\sqrt{\kappa}} \, \gtrsim 1. 6\times  10^7 \text{g} \,\gamma^{1/4} \left( \frac{ H_\text{end}}{10^{-6}\,M_\text{Pl}}\right)^{1/3} \left( \frac{g_*}{106.75}\right)^{1/12}\,.
\end{align}
We recall that the mass $M$ has to satisfy the upper bound given by the Eq. (\ref{Minter}), $M<M_\text{inter} \simeq 2 \kappa^{2/5} 10^6$g, otherwise it is always $f_\text{rem}<1$. This bound gives a maximum value for the $\kappa$, 
\begin{align}
\frac{\kappa}{10^{-10}} \lesssim 8.5 \, \gamma^{-5/2} \left( \frac{ H_\text{end}}{10^{-6}\,M_\text{Pl}}\right)^{-10/3}  \left( \frac{g_*}{106.75}\right)^{-5/6}\,.
\end{align}
Unless $H_\text{end} \ll 10^{-6}M_\text{Pl}$ it must be $\kappa <1$, hence for high scale inflation the  PBH remnants must have subplanckian masses.
For $\kappa=\kappa_\text{max}$ a maximum value for the mass of the PBHs, $M=M_\text{inter}$, is obtained for the kination regime in order the remnants to saturate the $\Omega_\text{DM}$.
%\begin{align}
%M_\text{max} =4.7 \times 10^2  \text{g} \, \gamma_\text{}^{-1} 
%\left( \frac{ H_\text{end}}{10^{-6}\,M_\text{Pl}}\right)^{-4/3}
 %\left( \frac{g_*}{106.75}\right)^{-1/3}\,.
%\end{align}

PBH remnants with $\kappa \geq  1$ require $H_\text{end} \lesssim 2 \times 10^{-9} \gamma_\text{}^{-3/4} M_\text{Pl}$, that can be achieved either in small field inflation model or by models where the CMB and PBHs potential energy scales have a large difference, % by placing the inflection point (or whatever generates a peak at the ${\cal P_R}(k)$) at small potential energy scales, 
so that the high frequency GWs  have a smaller amplitude.
We underline that the above results are valid only if the post-inflationary  equation of state of the inflaton field satisfies $w_\text{}\simeq 1$, at least until the BBN epoch. 
If it is $w_\text{}< 1$ the derived bounds get relaxed.

\renewcommand{\thesection}{4}
\section{Building a ${\cal P_R}(k)$ peak in accordance with observations}

\subsection{The position of the ${\cal P_R}(k)$ peak}
The wavenumber that inflation ends is 
\begin{align}
k_\text{end} = k_* \frac{H_\text{end}}{H_*} e^{N_*}
\end{align}
where $N_*$ are the e-folds of the observable inflation and given by the expression 
\begin{equation} \label{Nrh}
N_* \simeq 57.6 +\frac14 \ln \epsilon_* +\frac14 \ln \frac{V_*}{\rho_\text{end}} -\frac{1-3w_\text{}}{4} \tilde{N}_\text{rh} \,.
\end{equation}
%Let us call $\tilde{N}_\text{rh}$ the e-folds of the reheating stage 
%then the wavenumber at the moment of reheating completion is $k_\text{rh}=k_\text{end}e^{-\frac{3w+1}{2}\tilde{N}_\text{rh}}$. The $k_\text{end}$ is the wavenumber that inflation ends and it is related to the Planck pivot scale $k_*=0.05$ Mpc$^{-1}$ as $k_\text{end}=k_*e^{N_*} (H_\text{end}/H_*)$. The $N_*$ are the e-folds of the observable inflation,
The $\epsilon_*$, $H_*$, $V_*$ are respectively the first slow-roll parameter, the Hubble scale and the potential energy when the CMB pivot scale exits the Hubble radius, while $H_\text{end}$, $\rho_\text{end}$ are the Hubble scale and the energy density at the end of inflation.

The $N_*$ value is related to the postinflationary reheating efolds  $\tilde{N}_\text{rh}$ and the corresponding (averaged) equation of state ${w}_\text{}$. 
 We have implicitly assumed that $w$ refers to the postinflationary equation of state until the moment reheating  completes.  
 The number of efolds until the completion of the reheating $\tilde{N}_\text{rh}$ are
\begin{align}
\tilde{N}_\text{rh}&(T_\text{rh},H_\text{end}, g_*, {w}_\text{})=  \nonumber
\\
& -\frac{4}{3(1+ {w}_\text{})} \ln\left[\left(\frac{\pi^2 g_*}{90}  \right)^{1/4} \frac{T_\text{rh}}{(H_\text{end}M_\text{Pl})^{1/2}} \right]\,.
\end{align}
An inhomogeneity of size $k^{-1}$ crosses inside the horizon during radiation domination if $k<k_\text{rh}$ where
\begin{align}
k_\text{rh}=k_\text{end}e^{-\frac{3w+1}{2}\tilde{N}_\text{rh}}\,.
\end{align}
%and $k$ is the wavenumber that corresponds might lead to a PBH with mass $M$. 
For a general expansion rate determined by the effective equation of state value ${w}_\text{rh}$ the $k(M)$ relation reads,
\begin{align} \label{kMI}
k_\text{} (M, w_\text{})= k_\text{end}\,  \left(\frac{M/\gamma}{M_\text{end}}\right)^{-\frac{3w_\text{}+1}{3(1+w_\text{})}}\,.
\end{align}
%where $ M_\text{end}= {4\pi M^2_\text{Pl}}/{H_\text{end}}$ is the horizon mass at the end of inflation. 
The $M_\text{end}$, $k_\text{end}$ depend on the details of the inflationary model with the later becoming larger for larger values of the $w_\text{}$.
The $k(M, w_\text{})$ can be written using the reheating completion moment as the reference period replacing respectively the $k_\text{end}, M_\text{end}$ with the $k_\text{rh}, M_\text{rh}$ in Eq. (\ref{kMI}). Then we  attain the more general  $k(M, T_\text{rh}, w_\text{})$ relation
\begin{align} \label{kMgen}
k (M,  T_\text{rh}, &  w_\text{})  \simeq 2 \times 10^{17} \text{Mpc}^{-1} \left( \frac{T_\text{rh}}{10^{10}\text{GeV}} \right)^{\frac{1-3w}{3(1+w)}}  \nonumber \\
& \left(\frac{M/\gamma}{10^{12} \text{g}} \right)^{-\frac{3w+1}{3(1+w)}} \left(\frac{g_*}{106.75}\right)^{\frac{1}{4}\frac{1-3w}{3(1+w)}}
\end{align}
For the case of kination domination the minor correction, $T_\text{rh} \rightarrow 2^{1/4} T_\text{rh}$ should be added due to the equipartition of the energy density between the radiation and the scalar field.

Assuming a one-to-one correspondence between the scale of
perturbation and the mass of PBHs, an inflationary model builder who aims at generating PBHs with mass $M$ has to produce a ${\cal P_R}(k)$ peak at the wavenumber $k (M,  T_\text{rh},  w_\text{})$. Next we briefly discuss  the additional observational constraints, regarding the width of the peak, that one has to take into account in order the inflationary model to be viable. 

\begin{figure*}[htpb]
\centering
\includegraphics[width = 1.\textwidth]{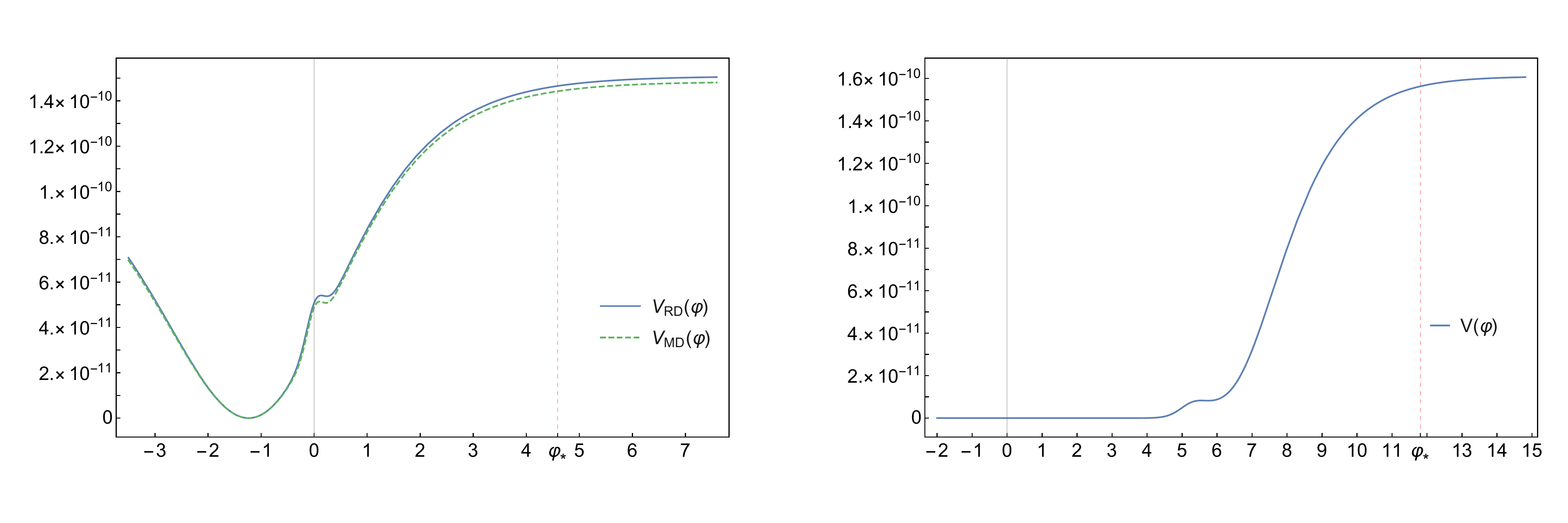}
\caption{ \textit{Left panel}: The potentials from the superconformal attractors, Eq. (\ref{VRM}), that trigger the PBH formation during the radiation era (solid line) and during matter era (dashed line).  Although the potentials differ slightly yield very different power spectra, see Fig. \ref{PSrad} and \ref{PSmat},  and PBH masses.
\textit{Right panel}: The potential for the runaway model, Eq. (\ref{Vrun}),  with the characteristic asymptotic flatness for large negative values of $\varphi$. 
In both panels the  position of inflection point and the total number of efolds $N_*$  determine the mass of the PBHs. The red dashed vertical line indicates the $\varphi_*$ position of the field that corresponds to the Planck pivot scale $k_*$; it is  $\varphi_*/M_\text{Pl}=4.59, 4.60, 11.81$, for the radiation, matter and kination cases respectively.
 The parameters of the potentials are listed in the Table III.
} \label{Vplots}
\end{figure*}

\subsection{Observational constraints on the ${\cal P_R}(k)$ at small scales}

A power spectrum peak at large wavenumbers, $k
\gg k_*$, is welcome for not spoiling the $n_s$ and $\alpha_s$ values measured at  $k_*$. 
Also, such a peak 
can generate PBHs abundant enough to comprise the entire dark matter in the universe, either as long lived PBHs or as PBHs remnants. 
However, shifting the peak at large wavenumbers does not render the ${\cal P_R}(k)$ free from constraints.
The impact of the Hawking radiation on the BBN and CMB observables and the extragalactic $\gamma$-ray background put strong upper bounds on the ${\cal P_R}(k)$ at large $k$-bands. These bounds  rule out a great part of the PBH mass spectrum with range $10^9 \text{g}<M< 10^{17}$g from explaining the $\Omega_\text{DM}$ with PBH remnants (PBHs remnants from holes with mass $M\sim 10^{10}$ g or larger could explain the   $\Omega_\text{DM}$  if $\kappa \gg 1$.) 
Moreover, even if the power spectrum peak produces PBHs with masses $M\sim 10^{5}$g,  $M\sim 10^{18}$g or $M\sim 10^{22}$g where the PBH abundance can be maximal, the width of the peak has to be particularly narrow.  The stringent constraint comes from the CMB, at the mass scale $M\sim 10^{13}$g, where the electrons and positron produced by PBHs evaporation after the time of recombination scatter off the CMB photons and heat the surrounding matter 
damping small-scale CMB anisotropies contrary to observations. 
The next stringent constraint applies at the mass range $M=10^{10}-10^{13}$ g that evaporate affects the BBN relics via hadrodissociation and photodissociation processes \cite{MacGibbon:1990zk, MacGibbon:1991tj, Kohri:1999ex, Carr:2009}.
% become important comes from the BBN observables 
%The electrons and positron scatter off the the CMB photons and heat the surrounding matter.  

 In Ref. \cite{Dalianis:2018ymb} the observational constraints have been explicitly translated into ${\cal P_R}(k)$ bounds.   
In a radiation dominated early universe, utilizing the Eq. (\ref{kMgen}) the CMB constraint for $M_\text{cmb} \equiv 2.5\times 10^{13}$g and $w_\text{rh} =1/3$ yields the bound 
\begin{align} 
 \label{kspacCMB}
\sigma\,(3 \times 10^{17}\, k_*) \, \lesssim \, 0.035 \, 
\left(\frac{\delta_c}{0.41}\right) \,.
\end{align}
The $\sigma$ is the variance of the comoving density contrast, $\sigma^2 \sim (4/9)^2   {\cal P_R}$ and the bound  ${\cal P_R}(4 \times 10^{17}\, k_*)\lesssim {\cal O}(10^{-3})$ is derived.

Turning to a matter dominated early universe  reheated at temperatures  $T_\text{rh} \lesssim 10^7\,\text{GeV}$ the variance of the density perturbations has to satisfy the CMB bound,
\begin{align}  \label{spinCMB}
&  \sigma  (k(M_\text{cmb}, T_\text{rh}))     \, \lesssim \, 
\text{Exp}\left[ \right.
-6.9- 0.09\, \ln \frac{T_\text{rh}}{\text{GeV}}  \\
 &+ 2\times10^{-3}\,\left(\ln\frac{T_\text{rh}}{\text{GeV}}\right)^2 
 \left. 
- 3\times10^{-5}\,
\left( \ln \frac{T_\text{rh}}{\text{GeV}}\right)^3 
\right]
 \nonumber
\end{align} 
 where $k(M_\text{cmb}, T_\text{rh})\simeq 3\times 10^{17}\,k_*\, \gamma_\text{}^{1/3}\left( T_\text{rh}/{10^{7}\, \text{GeV}}\right)^{1/3}$ $(g_*/106.75)^{1/12} $, according to the Eq. (\ref{kMgen}) for $w_\text{rh} =0 $.
During matter domination era  it is $\sigma \sim (2/5){\cal P_R}^{1/2}$ and for $\gamma_\text{} =0.1$,  $T_\text{rh}=10^7$ GeV  the constraint on the power spectrum reads ${\cal P_R}( 1.4 \times 10^{17}\,k_*)\lesssim {\cal O}(9\times 10^{-7})$. If the reheating temperature is $10^7 \text{GeV }\lesssim T_\text{rh} \lesssim 4\times10^8\,\text{GeV}$ then the BBN constraint on the power spectrum applies.
For $M_\text{bbn} \equiv 5\times 10^{10}$  and  $T_\text{rh}=10^8$ GeV the constraint reads ${\cal P_R}(10^{18}\,k_*)\lesssim {\cal O}(4\times 10^{-6})$, that is a bit weaker than the CMB. For larger reheating temperatures, $T_\text{rh} \gtrsim 10^9\,\text{GeV}$ the constraints get significantly relaxed.

For the case of kination domination the CMB constraint applies on the scale with wavenumber $k^{}(M_\text{cmb}, T_\text{rh}) $ and reads, 
\begin{align} \label{KB}
\sigma\,(k(M_\text{cmb}, T_\text{rh}) \, \lesssim \, 0.032 \, 
\left(\frac{\delta_c}{0.375}\right) \,.
\end{align}
It is $k(M_\text{cmb}, T_\text{rh}) \simeq 5 \times 10^{18}\, k_* \, \gamma_\text{}^{2/3}\left( T_\text{rh}/{10^{7}\, \text{GeV}}\right)^{-1/3}$ $(g_*/106.75)^{-1/12}$,  according to the Eq. (\ref{kMgen}) for $w_\text{rh}=1$ and $M_\text{cmb} =2.5\times 10^{13}$g. 
Note here that the CMB bound (\ref{KB}) applies for  reheating temperatures $T_\text{rh} \lesssim 2\times 10^{9}$ GeV since the gravitational collapse can be considered instantaneous, contrary to the case of the matter era, \cite{Dalianis:2018ymb}.

For smaller PBH masses that evaporate in less than a second 
there are limits on the amount of the thermal radiation from the PBHs evaporation due to the  production of entropy, that may be in conflict with the cosmological photon-to-baryon ratio, \cite{Zel'dovich3}. 
There are also  constraints from the abundance of dark matter produced by the evaporation, e.g. the lightest supersymmetric particle (LSP).
In our models the abundance of the PBH remnants saturates the dark matter density parameter $\Omega_\text{DM}h^2=0.12$, thus the LSP constraint does not apply.
Finally, for ultra small PBH masses, the constraint comes only from the relic abundance of the  Planck-mass remnants \cite{MacGibbon:1987my, Barrow, Carr:1994ar}.  These constraints are labeled {\it entropy, LSP} (with dotted-dashed line due to fact that in our models  this constraint is raised) and {\it Planck} respectively in our figures.

\renewcommand{\thesection}{5}
 \section{PBHs from the $\alpha$-attractors inflation models \label{PSP} }

\subsection{\label{potential} The inflaton potential and the computation of the  ${\cal P_R}(k)$}

The above constraints imply that, even in large wavenumbers, the power spectrum peak has to be positioned in a particular range of $k$ and, additionally, be sufficiently narrow.
In this Section our goal is to generate PBHs that will evaporate fast enough in the early universe without affecting the BBN and CMB observables and, at the same time,  leave behind mass remnants that will saturate the dark matter abundance. 
In order to implement this scenario we employ the machinery of $\alpha$-attractors and build inflationary models with inflection point at large $k$.

If the inflaton potential features an inflection point a large amplification in the power spectrum ${\cal P_R}(k)$ can be achieved due to the acceleration and deceleration of the inflaton field in the region around the inflection point as was pointed out in \cite{Garcia-Bellido:2017mdw}. The presence of an inflection point requires $V'\approx0$ and $V''=0$. 
In the context of supergravity such a model may arise from $\alpha$--attractors,  by choosing appropriate values for the parameters in the superpotential as described in Ref. \cite{Dalianis}.

We focus on the effective Lagrangian for the inflaton field $\varphi$ in the $\alpha$--attractors scenario that turns out to be 
\begin{equation} \label{la11}
e^{-1}{\cal L}= \frac{1}{2}R-\frac{1}{2}
\Big(\partial_\mu \varphi\Big)^2-f^2\Big(\tanh\frac{\varphi}{\sqrt{6\alpha}}\Big)\,,
\end{equation}
where Re$\Phi=\phi=\sqrt{3}\tanh({\varphi}/{\sqrt{6\alpha}})$ is a chiral superfield.

Polynomial and trigonometric forms for the function $f(\phi)$ can feature   an inflection point plateau sufficient to generate a significant dark matter abundance in accordance with the observational constraints  \cite{Dalianis}. 
Nevertheless, other forms for the function $f(\phi)$ are plausible. Exponential potentials enjoy a theoretical motivation in several BSM frameworks %, see e.g. \cite{Salam:1984cj, Nishino:1984gk, Maeda:1985es,Townsend:2001ea, Kehagias:2004bd} for some early works
 and their cosmology has been extensively studied, see e.g \cite{Copeland:1997et, Kolda:2001ex, Kehagias:2004bd, Russo:2004ym, Dalianis:2014nwa, Geng:2017mic, Basilakos:2019dof}.
 In the following we will examine the PBH formation scenario from $\alpha$-attractor inflationary potentials built by exponential functions.

The form of the potential fully determines the subsequent adiabatic evolution of the universe. Firstly, the number of efolds $N_*$, that follow the moment the $k_*^{-1}$ scale exits the quasi-de Sitter horizon, determine the duration of the non-thermal stage after inflation. Secondly, the position and the features of the inflection point plateau determine the mass and the abundance of the PBHs that form and, 
in particular, the moment the overdensities reenter the horizon. 
If PBHs of a given mass $M$ form during radiation era a specific inflationary potential has to be designed. On the contrary, PBHs production during matter era requires a different potential. Furthermore, an inflationary potential might be a runaway without minimum at all.
 Such a potential is acceptable if it can realize an inflationary exit and a sufficient reheating of the universe. Remarkably, both of these conditions can be satisfied in our modes with the  generated mini PBHs to  guarantee a successful reheating via their evaporation.
Last, but not least, at large scales $k\sim k_*$ we demand the CMB observables, as they are specified by the Planck 2018 data, to remain intact.

The PBH abundance is found only after the computation of the value of the comoving curvature perturbation ${\cal R}_k$. 
 In the comoving gauge we have  $\delta \varphi=0$ and $g_{ij}=a^2 \left[(1-2{\cal R}) \delta_{ij} +h_{ij}\right]$, Expanding the inflaton-gravity action
to second order in ${\cal R}$ one obtains
\begin{equation}
S_{(2)}= \frac12 \int {\rm d}^4x \sqrt{-g} a^3 \frac{\dot{\varphi}^2}{H^2} \left[\dot{{\cal R}}^2 -\frac{(\partial_i {\cal R})^2}{a^2}\right]\,.
\end{equation} 
After the variable redefinition $v=z{\cal R}$ where $z^2=a^2\dot{\phi}^2/H^2=2a^2\epsilon_1$ and switching to conformal time $\tau$ 
(defined by $d\tau=dt/a$), the action is recast into
\begin{equation}
S_{(2)} =\frac12 \int {\rm d}\tau {\rm d}^3x \left[(v')^2 -(\partial_i v)^2 +\frac{z''}{z}v^2 \right]\,.
\end{equation}
The evolution of the Fourier modes $v_k$ of $v(x)$ are described by the Mukhanov-Sasaki equation
\begin{equation} \label{MS}
v''_k +\left(k^2-\frac{z''}{z}\right) v_k =0 ,
\end{equation}
where $z''/z$ is  expressed in terms of the %Hubble flow 
functions  
\begin{equation} \label{eek}
\epsilon_1 \equiv -\frac{\dot{H}}{H^2}, \quad
 \epsilon_2 \equiv \frac{\dot{\epsilon}_1}{H\epsilon_1}, \quad 
 \epsilon_3 \equiv \frac{{\dot{\epsilon}}_2}{H\epsilon_2}, 
\end{equation}
as 
\begin{equation}
\frac{z''}{z} =(aH)^2 \left[2-\epsilon_1 +\frac32 \epsilon_2 - \frac12\epsilon_1 \epsilon_2 +\frac14 \epsilon^2_2+\frac{1}{2} \epsilon_2 \epsilon_3 \right]. 
\end{equation}
We are interested in the
the super-Hubble evolution of the curvature perturbation, that is for $k^2 \ll z''/z$.  
The power of ${\cal R}_k$ on a given scale   is obtained  once the  solution $v_k$ of the Mukhanov-Sasaki equation is known and estimated at a time well after it exits the horizon and its value freezes out,
\begin{equation}
 \left. {\cal P_ R} =\frac{k^3}{2\pi^2}\frac{|v_k|^2}{z^2} \right|_{k\ll aH } \,.  \label{ppp}
\end{equation}
After the numerical computation of the Mukhanov-Sasaki equation the  ${\cal P_ R}$  at all the  scales is obtained. 
As it is required, the ${\cal P_R}(k)$ of our models satisfy the constraints qiven by Eq. (\ref{kspacCMB}), (\ref{spinCMB}) and (\ref{KB}) for radiation, matter and kination eras respectively.
From the ${\cal P_ R}(k)$ we compute the $f_\text{rem}$  as described in Section 3. 
We note that we neglected possible impacts on the power spectrum from non-Gaussianities \cite{Franciolini:2018vbk, Atal:2018neu, DeLuca:2019qsy} and quantum diffusion effects \cite{Pattison:2017mbe, Biagetti:2018pjj, Ezquiaga:2018gbw, Cruces:2018cvq}.

\begin{figure}[htpb]
\centering
\includegraphics[width = 0.48\textwidth]{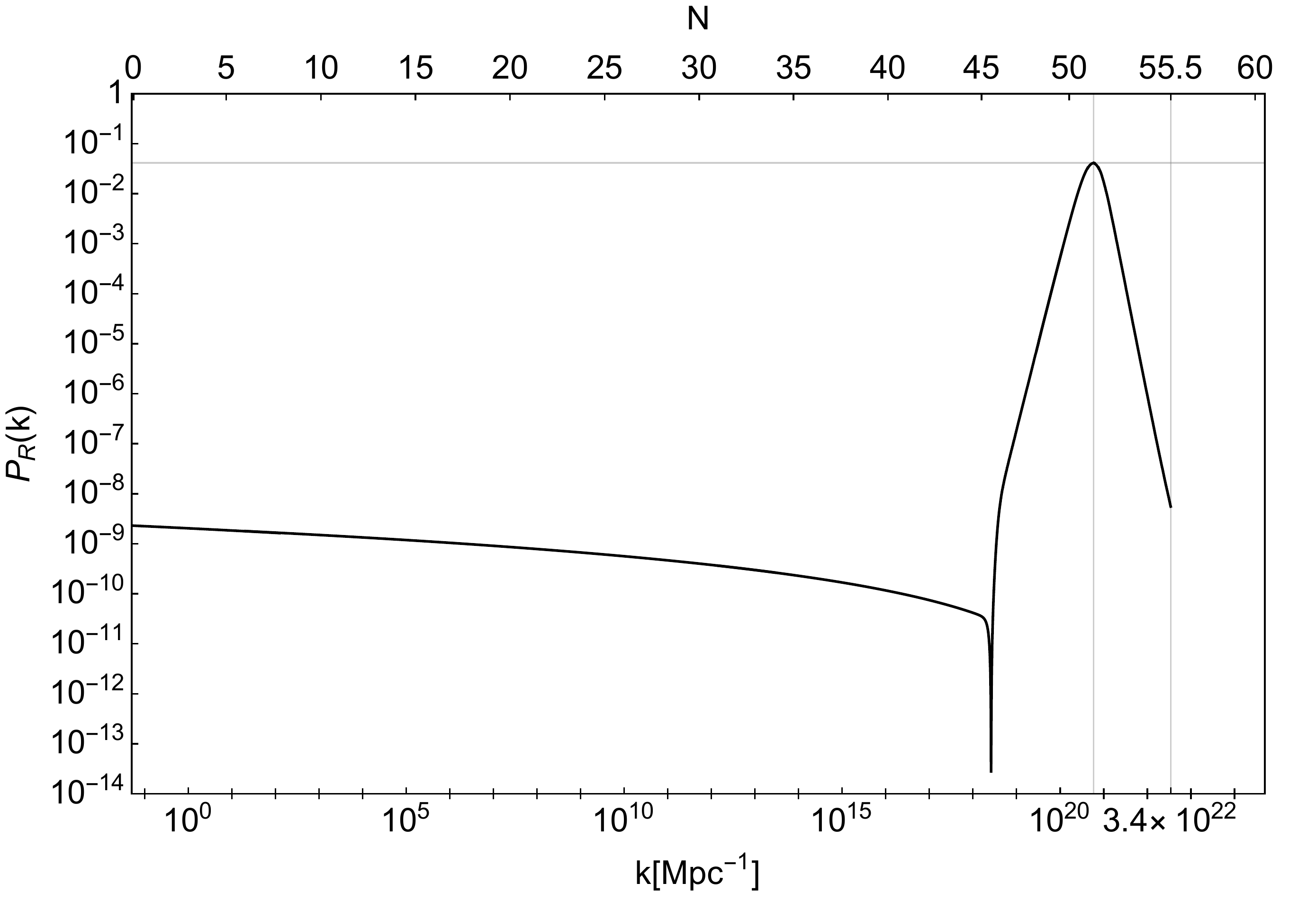}
\caption{The power spectrum of the comoving curvature perturbations for the 
model (\ref{VRM}) with parameters listed in the Table III for the RD case and potential depicted in the Fig. (\ref{Vplots}).
A significant amplification of the power spectrum $\mathcal{P}_{\mathcal{R}}\simeq 4\times10^{-2}$ takes place at small scales $k=5.9 \times 10^{20}$ Mpc$^{-1}$ about $N\simeq 4$ efolds before the end of inflation triggering the production of mini PBHs. The duration of the reheating era is almost instantaneous.
} \label{PSrad}
\end{figure}

\begin{figure}[htpb]
\centering
\includegraphics[width = 0.48\textwidth]{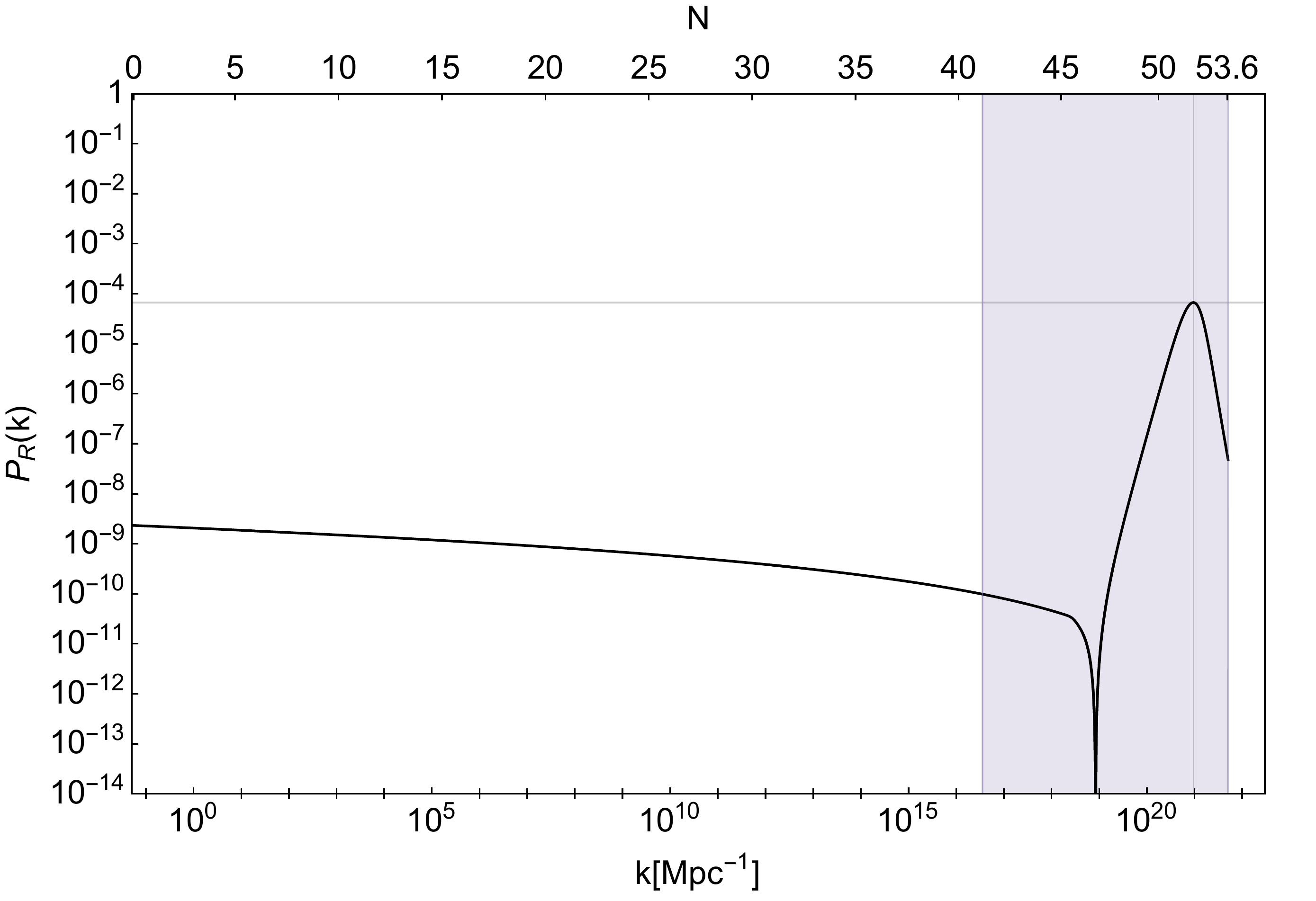}
\caption{
The power spectrum of the comoving curvature perturbations for the 
model (\ref{VRM}) with parameters listed in the Table III for the MD case and potential depicted in the Fig. (\ref{Vplots}).
The power spectrum has a peak with amplitude  $\mathcal{P}_{\mathcal{R}}\simeq 7\times10^{-5}$ at the scale $k=9.6 \times 10^{20}$ Mpc$^{-1}$ about $N\simeq 2$ efolds before the end of inflation and mini PBHs are produced. The shaded part  of the ${\cal P_R}(k)$ corresponds to the scales that enter the horizon   during the matter era.
} \label{PSmat}
\end{figure}

\begin{figure}[htpb]
\centering
\includegraphics[width = 0.48\textwidth]{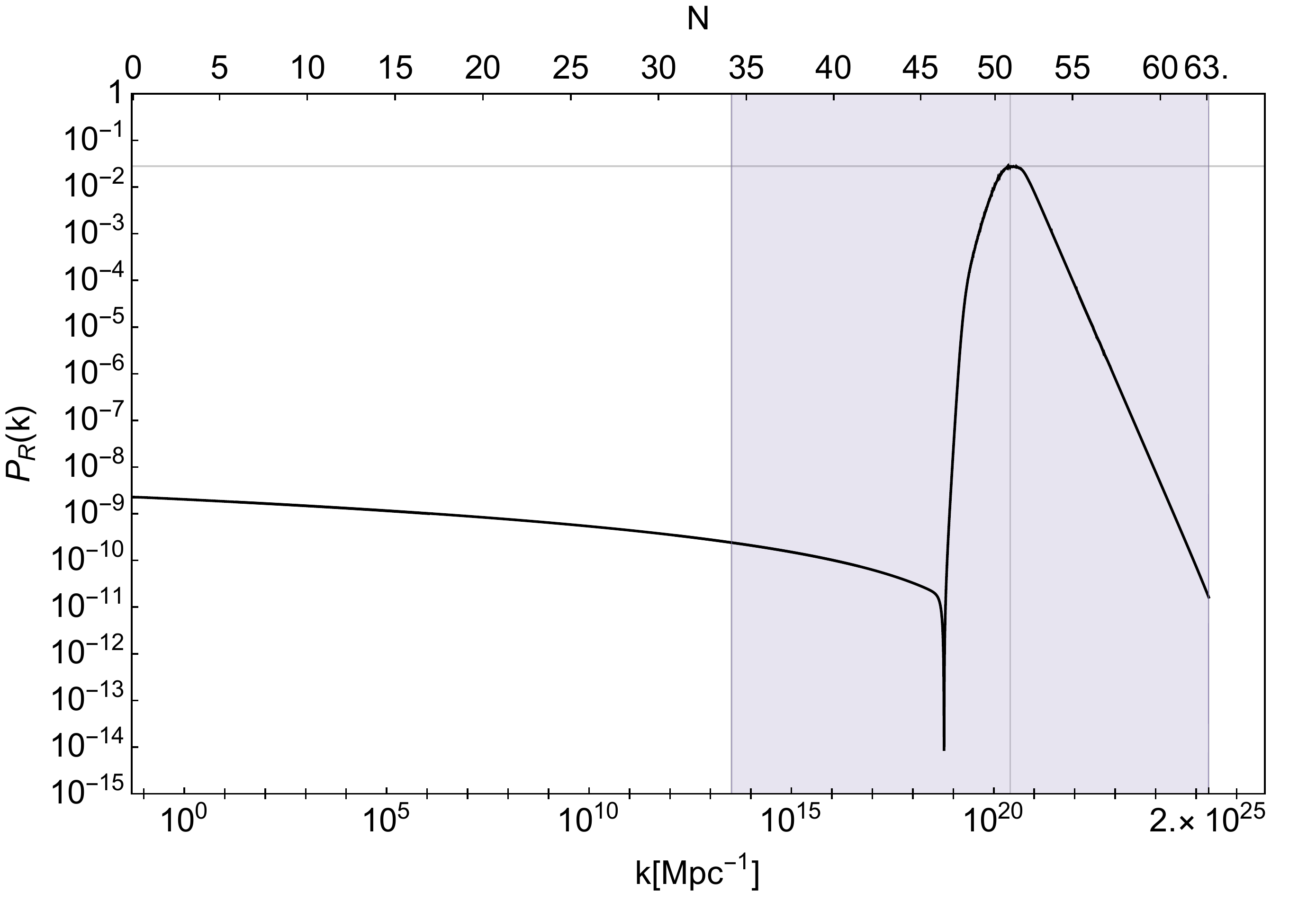}
\caption{
The power spectrum of the comoving curvature perturbations for the 
model (\ref{Vrun}) with parameters listed in the Table III for the SD case and potential depicted in the Fig. (\ref{Vplots}).
The power spectrum has a peak with amplitude $\mathcal{P}_{\mathcal{R}}\simeq 2.7  \times10^{-2}$  at the scales  $k=2 \times 10^{25}$ Mpc$^{-1}$ about $N\simeq 12$ efolds before the end of inflation triggering the production of mini PBHs during kination regime. The shaded part of the ${\cal P_R}(k)$ corresponds to the scales that enter the horizon  during  the  kination era.
} \label{PSkin}
\end{figure}

\subsection{Inflaton potential for PBHs production during radiation and matter era}

\subsubsection{Radiation era}

A function $f(\phi)$ built by exponentials 
can feature a proper inflection point plateau. 
The form of the potential is chosen to produce PBHs of the right abundance.
We ask for large reheating temperatures so that the large inhomogeneities reenter the horizon after the thermalization of the universe.
This is achieved by sufficiently strong couplings of the  inflaton field to the visible sector. 
We also demand values for the $n_s$ and $\alpha_s$ that are favorable by Planck 2018 data \cite{Akrami:2018odb}. 
An example of a combination of exponentials that can fulfill the above requirements is of the form, 
\begin{align} \label{fphi}
f(\phi/ \sqrt{3}) =f_0\, ( c_0 + c_1 e^{\lambda_1 \phi/\sqrt{3}} + c_2 e^{\lambda_2 \phi^2/3})
\end{align}
 that generates the potential 
\begin{align} \label{VRM}
V(\varphi)= f^2_0\left( c_0 + c_1 e^{\lambda_1 \tanh\varphi/\sqrt{6}} + c_2 e^{\lambda_2 (\tanh \varphi/\sqrt{6})^2} \right)^2
\end{align}
having taken $\alpha=1$.
The determination of the parameters values requires a subtle numerical process  that we outline.
Firstly, a central PBH mass  $M$ has to be chosen and from Eq. (\ref{fRD2}) the  $\beta$ value that saturates the $f_\text{PBH}$  is specified. 
The $M$ is the parameter that spots the $k$-position of the ${\cal P_R}(k)$ peak. The $\beta$  is exponentially sensitive to the  amplitude of the 
 peak and its exact value is found after a delicate selection of the potential parameters.
 The ${\cal P_R}(k)$ is produced by solving numerically the Mukhanov-Sasaki equation, following the method described in Ref. \cite{Dalianis}.    
 At  the same time consistency with the CMB normalization, the measured $n_s$ and $\alpha_s$ values, as well as large enough $N_*$ values are required. 
 %A slight change in the parameters results in a different cosmology.  

For $\phi \rightarrow \sqrt{3}$ the potential drives the early universe cosmic inflation and the CMB normalization gives the first constraint for the parameters.  We also demand zero potential energy at the minimum of the potential that gives a second constraint. We also note that the $f_0$ is a redundant parameter since it can be absorbed by the  $c_0$, $c_1$ and $c_2$. We keep only for numeric convenience.
  An example of parameter values that realize the PBH production during RD era is listed in the Table III and the potential is depicted in the left panel of the Figure {\ref{Vplots}}.

 \subsubsection{Matter era}

An early universe matter domination era can be realized if the shape of the inflationary potential around the minimum is approximated with a quadratic potential. For moderately suppressed inflaton couplings the inflaton decays after a large number of oscillations. 
% Its energy density during oscillations acts effectively as pressureless matter and the perturbative decay reheats the universe with a temperature $T_\text{rh}<10^{15}$ GeV. 
The inhomogeneities that reenter the horizon during the stage of the inflaton oscillations might collapse in a pressureless environment.

%Therefore, the inflationary potential has to have a 
The calculation of the PBH production during matter era involves the same numerical steps with the case of radiation plus some  extra conditions that have to be taken into account. 
Firstly, the PBH mass $M$ value is not adequate to specify the $k$-position and the amplitude of the ${\cal P_R}(k)$ peak, since there is a crucial dependence on the reheating temperature. Hence, after choosing the PBH  mass $M$, the required $\beta$ is fixed for a particular reheating temperature. In turn, the $T_\text{rh}$ fixes the number of efolds that constrain the inflaton excursion in the field space.
Moreover, the amplitude of the peak has an additional dependence on the reheating temperature, namely the variance of the perturbations has to satisfy the bound (\ref{sigmaCr}), $\sigma>\sigma_\text{cr}(T_\text{rh})$, in order the inhomogeneities to fully collapse during the matter domination era.

An inflationary example that predicts PBH formation during matter era is given by Eq. (\ref{fphi}) after proper parameter values are chosen.  
The  set of the parameters, listed in Table III, yields an amplitude for the ${\cal P_R}(k)$  that spin effects have to be considered in the estimation of the formation probability.

\begin{center}\label{tab2}
 \begin{tabular}{|c||c|c|c|c|c|} 
\hline
  $\textbf{\textit{era}}$  
  & $\boldsymbol{\beta}$ 
  & $\boldsymbol{T_\text{rh}}$ (GeV)
  & $\boldsymbol{\tilde{N}_\text{rh}}$
  & $\boldsymbol{M_\text{peak} (\text{g})}$ 
  & $\boldsymbol{f_\text{rem}}$  
    \\ 
 \hline \hline
\textbf{RD}& 1.25$\times 10^{-13}$ & $2.13\times 10^{15}$& 0 &  3.9 $\times 10^{4}$ & $\sim 1$
%& RD 
\\ 
 \hline
  $\textbf{MD}$ & 6.88$\times 10^{-16}$ & $5.6\times 10^{12}$ & 7.9 &  61 & $\sim 1$ \\
%  
% \hline
%   $\textbf{MD}_2$ & 6.88$\times 10^{-16}$ & 2.7$\times10^{11}$& 11.9 &  $\sim$ 61$\times 10^{0}$ & $<$1
%   \\
   \hline
      \textbf{KIN}& 1.07$\times 10^{-14}$ & $1.8\times 10^{6}$& 13.57&  $2 \times 10^{2}$ & $\sim 1$
   \\
   \hline
%& KIN 
\end{tabular}
\captionof{table}{\label{tabRad}
The predictions of the three inflationary models discussed in the text. The era indicates the production era of the PBHs, that is the era that the  perturbations with the largest amplitude re-enter the horizon.
 %Example of values for the production of the PBHs, the remnants and the reheating epoch are presented. 
 The PBH remnant abundance,  $f_\text{rem}$ is the maximal one. %$\boldsymbol{f_\text{rem}}$ for 
Depending on the parameters of the potentials the initial PBH mass $\boldsymbol{M^\text{peak}_\text{PBH} }$ varies in the range $M \sim 10-10^4 \text{g}$. 
}
\end{center}

 \begin{center}\label{potpar1}
\begin{tabular}{|c||c|c|c|} 
\hline
 $\textbf{\textit{era}}$  & $\boldsymbol{{\cal P_R}^\text{peak}}$ & $k_{\text{end}}$ & $\boldsymbol{N_*}$ \\ [0.5ex] 
 \hline \hline
\textbf{RD} &   4$\times 10^{-2}$ & 3.45$\times10^{22}$ & 55.5  \\ 
\hline
$\textbf{MD}$ & 6.6 $\times 10^{-5}$ & 5.06$\times10^{21}$ & 53.6 \\ 
\hline
\textbf{KIN} &  2.7$\times10^{-2}$ & 2.$\times10^{25}$ & 63.    \\ 
\hline
\end{tabular}
\captionof{table}{
Characteristic values for the curvature power spectrum are listed. 
} 
\end{center}
% 7.49267$\times10^{-11}$ & 7.37421$\times10^{-11}$ & 3.11497$\times10^{-62}$ &  & $\boldsymbol{V_0}$
\begin{center}\label{potpar2}
\begin{tabular}{|c|c|c|c|c|c|c|} 
\hline
 $\textbf{\textit{era}}$ & $ \boldsymbol{c_0}$ & $ \boldsymbol{c_1}$ & $ \boldsymbol{c_2}$ & $\boldsymbol{\lambda_1}$ & $\boldsymbol{\lambda_2}$ 
 %& $\boldsymbol{\varphi_*}$ 
 \\ [0.5ex] 
 \hline \hline
\textbf{RD} & -1.856  & 1.173 & -0.14 & -0.987 & 95.5904 \\ %& 4.5905 \\ 
\hline
\textbf{MD} & -1.856 & 1.173 & -0.13 & -0.987 & 124.555 \\ % & 4.60\\ 
%\hline
%\textbf{MD}& -1.856 & 1.173 & -0.13 & -0.987 & 124.546 & 4.578\\ 
\hline
\textbf{KIN}& $-8.70\times 10^{-27} $ & 0.1045 & $-4\times 10^{25}$ & 62.2 & -4430.973 \\ %& 11.812 \\ 
\hline
\end{tabular}
\captionof{table}{
A set of values for the parameters of each potential Eq. (\ref{VRM}), (\ref{Vrun})  responsible for PBH production in radiation, matter and kination domination scenarios are listed.
For the kination model we also took $\varphi_\text{P}=0.995 M_\text{Pl}$. 
We add that  $f^2_0=  7.49267 \times10^{-11}, \,  7.37421\times10^{-11},\,3.11497 \times10^{-62}$  for the radiation, matter and kination cases respectively.
 %The $\varphi_*$ is the value of the inflaton field when the scale $k_*=0.05\text{Mpc}^{-1}$ exits the horizon $N_*$ e-folds before the end of inflation. 
} 
\end{center}

\subsection{Inflaton potential for PBHs production during kination era}

A period of kination domination  has an interesting and distinct cosmology. It is possible to be realized 
 after inflation if the  potential  does not have a vacuum, see \cite{Dimopoulos:2017zvq, Dimopoulos:2017tud} for $\alpha$-attractor kination models. 
 A non-oscillatory inflaton field will runaway without decaying  resulting in a period where the kinetic energy dominates over the potential energy. 
The attractive feature of such models is that the inflaton can survive until today and 
might play the role of quintessence.
Moreover, such models are attractive because they lead to a different early universe phenomenology since the effective equation of state is $w_\text{}\sim 1$ and the expansion rate is reduced. This is the so-called stiff fluid or kination era that gives rise to different prediction regarding some early universe observables such as the spectrum of the tensor perturbations,  a fact that renders such an era testable.

The kination scenarios usually suffer from radiation  shortage since the inflaton field does not decay and special mechanisms have to be introduced.
A source of radiation comes from the Hawking temperature of de Sitter space, called gravitational reheating, but this is very inefficient \cite{Ford:1986sy, Chun:2009yu}.  On the other hand, the Hawking radiation from mini PBHs formed by a runaway inflaton  automatically reheat the universe.
So, in our models radiation is produced by the evaporation of the PBHs that can be efficient enough.  According to Eq. (\ref{Tkination}),
common values for the $\beta$ imply large enough reheating temperatures. 

The construction of kination inflation models that induce the PBH production is very challenging. 
Firstly, the inflaton runs away until it freezes at some value $\varphi_F$ and this residual potential energy of the inflaton must not to spoil the early and late time cosmology.  
The inflaton potential energy at $\varphi_F$  has to be tuned to values $V(\varphi_F) \lesssim 10^{-120} M^4_\text{Pl}$, similarly to all the quintessence models.
Secondly, the kination inflaton model parameters are self-constrained.
A particular PBH mass $M$  specifies the $k$ of the ${\cal P_R}(k)$
only if the reheating temperature is known.
 However, the reheating temperature is not a free parameter, as e.g in matter or radiation cases where the $T_\text{rh}$ depends on the inflaton decay rate. 
In the kination scenario the $T_\text{rh}$ depends on the $\beta$. 
The $\beta$ is found by the condition $f_\text{rem}=1$ and this fixes the reheating temperature. %, that also has to satisfy the constraint.. 

Hence, the characteristics of the  peak in the power spectrum determines
\begin{itemize}
\item [i)] the mass of the evaporating PBHs,
\item [ii)]  the dark matter abundance, and
\item [iii)] the reheating temperature of the universe.
\end{itemize}
% i) , ii) and iii) t.
In addition,  the tail of the potential might lead to the observed late time acceleration of the universe. Undoubtedly, this scenario is remarkably economic.

To be explicit, let us introduce the model 
\begin{align} \label{runf}
f(\phi/\sqrt{3})=f_0\,\left(c_0+  c_1 e^{\lambda_1 \phi/\sqrt{3}} + c_2 e^{\lambda_2 (\phi-\phi_\text{P})^2/3} \right) 
\end{align}
that generates the potential % $V(\varphi)=f^2\left(\tanh(\varphi/\sqrt{6})\right)$,
\begin{align} \label{Vrun}  \nonumber
 V(\varphi)=  f^2_0  \, \big[ c_0 + &  c_1 e^{\lambda_1 \tanh \varphi/\sqrt{6}} \, + 
 \\
&  c_2   e^{\lambda_2 \left(\tanh(\varphi/\sqrt{6})-\tanh(\varphi_\text{P} /\sqrt{6})\right)} \big]^2\,.
\end{align}
The $\varphi_\text{P}$ is a fixed value in the field space that determines the position of the inflection point. Again here the $f_0$ can be absorbed in $c_0$, $c_1$ and $c_2$.
For $\phi \rightarrow \sqrt{3}$  the early universe cosmic inflation takes place and the CMB normalization gives the first constraint for the parameters. 
 For $\phi \rightarrow -\sqrt{3}$ we demand zero potential energy, thus we get the second constraint
\begin{align} \label{c0}
c_0=-c_1 e^{-\lambda_1} - c_2 e^{\lambda_2 (\sqrt{3}+\phi_\text{P}))^2/3}\,.
\end{align}
 The kination stage  lasts until the moment that the radiation produced by the PBH evaporation dominates the energy density.  
Later the field freezes at some value $\phi_F$ and defreezes at the present universe.
The runaway potential is flat enough to lead to the currently observed accelerated expansion, hence implement a wCDM cosmology as a quintessence model.

\begin{figure}[htpb]
\centering
\includegraphics[width = 0.48\textwidth]{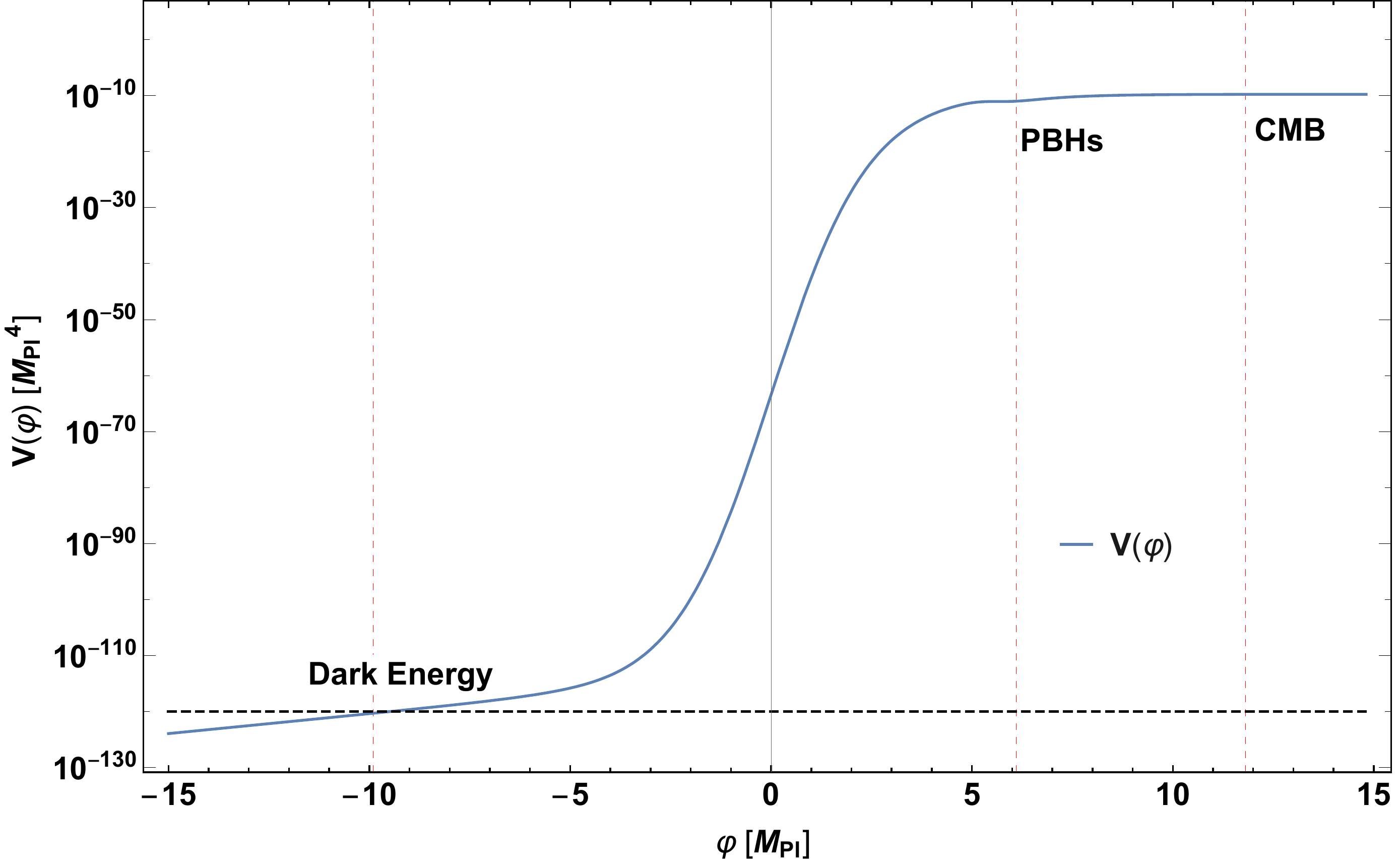}
\caption{
The potential (\ref{Vrun}), depicted in Fig. \ref{Vplots}, for the parameters listed in Table III (KIN). The  plot is in logarithmic scale in order to make  the $V(\varphi)$ value visible both during inflation and today. See also Fig. \ref{Fig1Vkin}. 
} \label{FigLogKin}
\end{figure}

Let us pursue some approximate analytic expressions that describe  the post-inflationary evolution of the field $\varphi$. After inflation the $\varphi$ rolls fast the potential and a stage of kination commends, where $\dot{\varphi}^2/2 \gg V(\varphi)$.
The Klein-Gordon equation for the $\varphi$ for negligible potential energy is $\ddot{\varphi}+3H\dot{\varphi}\simeq 0$. 
During kination it is $a\propto t^{1/3}$
and for $t\gg t_\text{end}$ the field value evolves as 
\begin{align}
\varphi-\varphi_\text{end} \simeq -\sqrt{\frac{2}{3}} M_\text{Pl} \ln\left( \frac{t}{t_\text{end}}\right)\,
\end{align} 
where we considered negative initial velocity for the $\varphi$.
At the moment $t_\text{form}$ the PBHs form and later at $t_\text{evap}$ they evaporate. Later, at the moment $t_\text{rh}$ the universe becomes radiation dominated and the kination regime ends.
Until reheating it is $a \propto t^{1/3}$ and one finds that $t_\text{rh}=(\Omega_\text{rad}(t_\text{evap}))^{-3/2} t_\text{evap}$, 
where  $\Omega_\text{rad}(t_\text{evap})=(3/2)\gamma^2_\text{} \beta M^2/m^2_\text{Pl}$, given by Eq. (\ref{KinEvap}).
At the moment of reheating the field value, $\varphi_\text{rh}$, is
\begin{align}
\varphi_\text{rh}\simeq \varphi_\text{end}-\sqrt{\frac{2}{3}} \left(-\frac{3}{2} \ln \Omega_\text{rad}(t_\text{evap}) + \ln\left(\frac{t_\text{evap}}{t_\text{end}}\right)\right)M_\text{Pl}
\end{align}
After reheating it is $a \propto t^{1/2}$ and the field evolution slows down,
\begin{align}
\varphi-\varphi_\text{rh} \simeq -\frac{2}{\sqrt{3}} M_\text{Pl} \left(1-\sqrt{ \frac{t_\text{rh}}{t}}\right)\,.
\end{align} 
For $t \gg t_\text{rh}$ the field gets displaced $2 M_\text{Pl}/\sqrt{3}$ from $\varphi_\text{rh}$
and thus, at some late moment $t_F$ 
the field freezes at the value $\varphi_F$, 
\begin{align}
\varphi_{F}\simeq \varphi_\text{end}-\sqrt{\frac{2}{3}} \left(\sqrt{2}-\frac{3}{2} \ln \Omega_\text{rad}(t_\text{evap}) + \ln\left(\frac{t_\text{evap}}{t_\text{end}}\right)\right)M_\text{Pl}
\end{align}

Asking for $f_\text{rem}=1$ we 
find  from Eq. (\ref{fremKIN}) that $ \Omega_\text{rad}(t_\text{evap})=3\times 10^{-13} (M/10^5 \text{g})^{10} (4\kappa)^{-4}$. Also it is $t_\text{evap} \sim 4\times 10^2(M/10^{10} \text{g})^3$ s and $t_\text{end} \simeq t_\text{Pl}(m_\text{Pl}/H_\text{end})$. Therefore we
obtain an expression for the $\varphi_F$ that depends only on the initial mass of the PBH $M$ and the mass of the PBH remnant $\kappa m_\text{Pl}$,
\begin{align} \label{fF}
\varphi_{F}\simeq \varphi_\text{end} - \sqrt{\frac{2}{3}} 
\left[19 +13 \ln(M/10^5 \text{g})+4 \ln(1/\kappa) \right]M_\text{Pl}\,.
\end{align} 
This is a general approximate expression for any runaway potential that predicts PBH remnants as dark matter. It is general because we have omitted the potential $V(\varphi)$ both from the Friedman and  the Klein-Gordon equations as negligible.
The  $\varphi_F$ value depends only on the mass $M$ and the parameter $\kappa$. %not on the $\beta$. 
For $\kappa=1$ and $M=10^5$g  it is $\varphi_F -\varphi_\text{end} \sim -15M_\text{Pl}$.  
For $\kappa=10^{-10}$ and $M=10^2$g  it is $\varphi_F -\varphi_\text{end} \sim -17M_\text{Pl}$. 
We note that the exact value of the $\varphi_F$ is found after the numerical solution of the Klein-Gordon and Friedman equations and the $|\varphi_F -\varphi_\text{end}| $ is a bit less than the value of the Eq. (\ref{fF}) for we neglected the potential $\varphi$ and considered instant transitions between the kination and radiation regime.

If we want to identify the dark energy as the energy density of the scalar field $\varphi$ then we have to tune the potential energy value at $\varphi_F$. 
For our model (\ref{Vrun}) we impose the condition, 
\begin{align} \label{third}
\frac{\rho_\text{inf}}{\rho_0} \simeq \frac{V(\varphi\gg 1)}{V(\varphi_F)}\sim \frac{e^{2\lambda_1}}{e^{-2\lambda_1}} \sim 10^{108}\,,
\end{align}
dictated by the hierarchy of energy scales between the $\alpha$-attractors inflation and the dark energy. This condition gives a third constraint to the parameters of the potential, together with  the CMB normalization  and the requirement for zero vacuum energy as $\varphi \rightarrow -\infty$, Eq. (\ref{c0}). The Eq.  (\ref{third}) 
gives a rough relation for the size of the exponent parameter $\lambda_1$, 
\begin{align}
4\lambda_1 \sim 108\ln(10)\,.
\end{align}

In the Table III we list 
  a set of parameters that the kination model (\ref{Vrun}) generates PBHs which after evaporation leave behind remnants with $f_\text{rem}=1$ and acts as quintessence.

\begin{figure}[htpb]
\centering
\includegraphics[width = 0.48\textwidth]{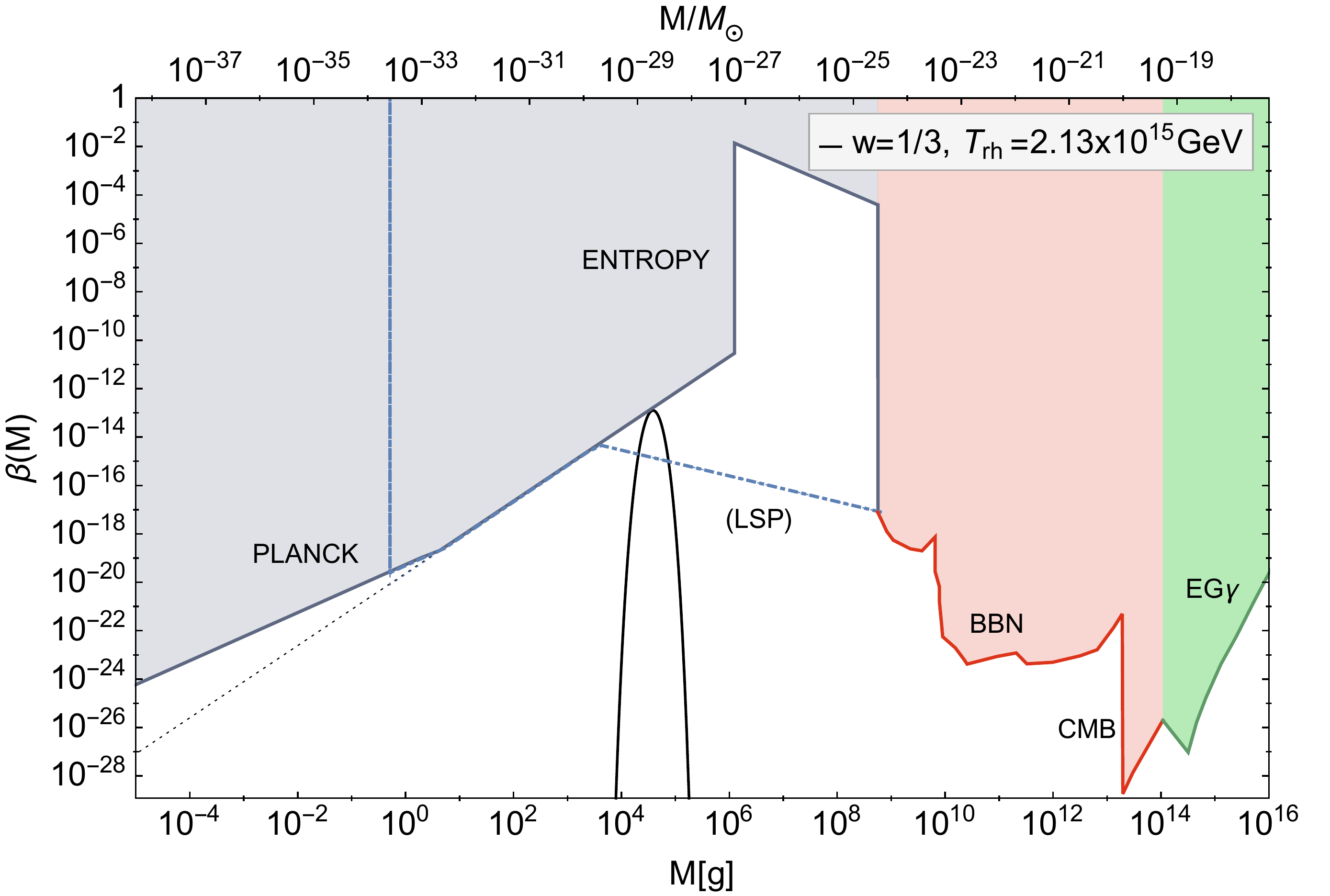}
\caption{
The PBH formation rate $\beta(M)$ estimated by the  Eq. (\ref{brad}) for the model (\ref{VRM}) with the parameters listed in the table III and ${\cal P_R}(k)$ depicted in Fig. \ref{PSrad} for the RD case. 
The central mass of the PBHs is $M \simeq 4 \times10^4$g that evaporate leaving behind Planck mass remnants with $f_\text{rem}=1$. The dotted black lines depicts the constraints if the reheating temperature was $T_{\text{rh}}>10^{15}$ GeV. The LSP upper bound is not applicable since the PBH remnants  comprise the total dark matter in our scenarios.} \label{FigbetaRD}
\end{figure}

\begin{figure}[htpb]
\centering
\includegraphics[width = 0.48\textwidth]{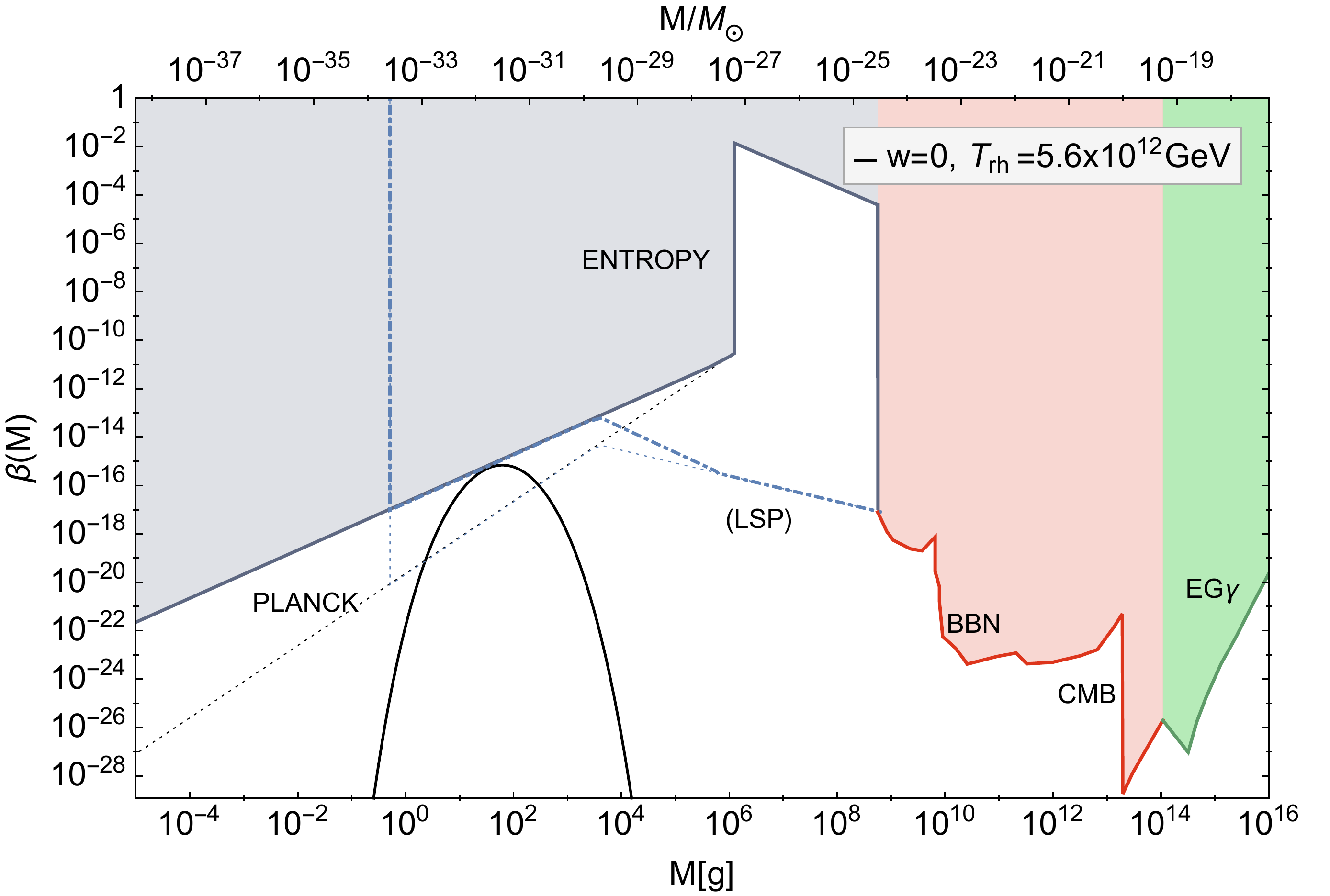}
\caption{
The PBH formation rate $\beta(M)$ estimated by the Eq. (\ref{bspin}) for the model (\ref{VRM}) with the parameters listed in the table III and ${\cal P_R}(k)$ depicted in Fig. \ref{PSmat} for the MD case. 
The central mass of the PBHs is $M \simeq 61$g that evaporate leaving behind Planck mass remnants with $f_\text{rem}=1$.
The dotted black and blue lines depict the constraints for arbitrary large reheating temperature. 
} \label{FigbetaMspin}
\end{figure}

 \subsection{CMB observables }

%[The dynamics of such a potential need a delicate approach while calculating the power spectrum of the quantum fluctuations, since the analytic approximation might lead to important underestimation of the quantities \cite{Ballesteros:2017fsr} due to the failure of slow-roll approximation (SR) near the shallow minimum]

The $n_s$ and $r$ values in the standard $\alpha$-attractors are expressed as the analytic relations, $n_s\sim1-2/N_*$ and $r\sim12\alpha/N_*^2$. 
These expressions still apply in $\alpha$-attractor models that feature an inflection point, with the essential difference that the $N_*$ is replaced by the number of efolds $\Delta N$ that separate the moments of horizon exit of the  CMB scale $k^{-1}_*$  and the PBH scale $k^{-1}_\text{}$. Thus it is $n_s\sim1-2/\Delta N$ and $r\sim12\alpha/\Delta N^2$.
In our models we get $\Delta N \gtrsim 50$  
hence the spectral index value is predicted to be 
\begin{align}
n_s \gtrsim 0.96
\end{align}
and the tensor-to-scalar ratio
\begin{align}
r< 0.048
\end{align}
placing the prediction of our models in the  68\% CL region of the Planck 2018 data \cite{Akrami:2018odb} % $n_s= 0.9649\pm0.0042$ and $r< 0.070$, 
without assuming running of the running for the $n_s$.
Generally, the $n_s$ value becomes larger than 0.96 if the PBHs have mass less than about $10^5$ grams.

%In this analysis we have studied the possibility of producing primordial black hole relics from $\alpha$-attractors while being compatible with the attractor scenario model prediction as well as with the observational CMB predictions...

\renewcommand{\thesection}{6}
\section{\label{concsection} Conclusions}

In this work we investigated the cosmology of mini  primordial black holes. 
The very motivation of examining this scenario is the theoretical postulation that a stable or long lived remnant is left behind after the evaporation of the "black" holes.
 The mass of the remnant is expected to depend on the unknown physics that operates at the Planck energy scale. Therefore we examined the cosmology of PBH remnants with arbitrary mass $M_\text{rem}=\kappa \, m_\text{Pl}$ and $\kappa$ a free parameter that might be orders of magnitude larger or smaller than one. 
The PBH remnants can comprise the entire dark matter of the universe if the mass of the parent PBH is roughly $M \lesssim  \, \kappa^{2/5} 10^6$ g. 
 %PBH mass has to be less than  less than about $10^8$ grams and 
 %hence the evaporate promptly in the early universe safely before the era of the big bang nucleosynthesis. 
We computed the general relic abundance of the PBHs remnants and 
found the conditions that they  comprise the entire cold dark matter in the universe.
We found that the PBH remnants have a significant  cosmological abundance only if they have mass $M_\text{rem}> 1$ GeV. 
Also the mass of the remnants  must have mass $M_\text{rem} \ll 10^8$ grams; otherwise the parent PBH affects the BBN or the CMB observables.

Mini PBHs imply that the comoving curvature perturbation is enhanced  at the extreme end of the ${\cal P_R}(k)$.
This is a rather attractive feature since the required large primordial inhomogeneities can be produced by the inflationary phase without spoiling the spectral index value $n_s$. 
% and inflationary models in accordance with the CMB constraints can be built. %In addition 
The PBHs form in the very early universe after the inflationary phase, hence the primordial inhomogeneities are expected to collapse during a non thermal phase unless the inflaton field decays very fast.

In this work we built inflationary models in the framework of $\alpha$-attractors. 
We produced a peak in power spectrum by constructing an inflection point and computed the  numerically the ${\cal P_R}(k)$   by solving the Mukhanov-Sasaki equation. Our models yield a spectral index value $n_s>0.96$, that places them in the $68\%$ CL contour region of Planck 2018 data.  
 The building blocks of the inflationary potentials are exponential functions. 
We examined the PBHs production for three different inflationary scenarios. 
In the first, the inflaton field decays nearly instantaneously after  inflation reheating the universe at very large temperatures. In this scenario the mini PBHs are produced and evaporate during the radiation phase. In the second scenario  the inflaton field decays a bit later, after oscillating several times about the minimum of its potential resulting in a post-inflationary stage of pressureless matter domination. During matter domination the primordial inhomogeneities collapse into PBHs. After the inflaton decay  the universe is reheated and the mini PBHs evaporate. 

In the third scenario the PBH are produced during a kination regime. This is a novel scenario, hence we examined it in more detail.
A kination regime takes place if the inflaton potential has no minimum and the inflaton runs away after the end of inflation. 
%In such a scenario the energy density of the mini PBHs  redshifts much slower than the background. %increases fast compared to the dominant inflaton field kinetic energy. % of the   until the moment of their complete evaporation. T
The radiation is produced by the PBH evaporation that gradually dominates the energy density and reheats  the universe. The resulting reheating temperature can be larger than $10^6$ GeV terminating fast enough the kination era in accordance with the BBN constraints.  The PBHs remnants %, that have mass less than $m_\text{Pl}$ for high scale inflation, 
can account for the entire dark matter of the universe. Interestingly enough,  the non-decaying inflaton can additionally act as quintessence field giving rise to the observed late time accelerated expansion implementing a wCDM cosmological model.  Actually this model is remarkably  economic in terms of ingredients.

Nowadays, that the existence of black holes and  dark matter are unambiguous,  the investigation of the PBH dark matter scenario is very motivated. Here we examined the less studied mini PBH scenario and derived general expression complementing older results and put forward new and testable cosmological scenarios for the early and late universe.
%Finally, let us comment that 
%We derived analytic expressions and plotted our results. 

\section*{Acknowledgments}

\noindent 
The work of I.D. is supported by the IKY Scholarship Programs for Strengthening Post Doctoral Research, co-financed by the European Social Fund ESF and the Greek government.

\noindent

%\pagebreak

\end{document}